\documentclass[12pt, a4paper]{article}

\usepackage[margin=1.in]{geometry}
\usepackage[compress,numbers,sort]{natbib}
\usepackage{xcolor}
\usepackage{comment}
\pdfoutput=1
\usepackage{graphicx,array}
\usepackage[T1]{fontenc}
\usepackage{amssymb,slashed,latexsym}
\usepackage{bm}
\usepackage{braket}
\usepackage{placeins}
\usepackage{adjustbox}
\usepackage{enumitem}
\usepackage{slashed}
\usepackage{color, verbatim}
\usepackage{latexsym}
\usepackage{amsmath}
\usepackage{graphicx}
\usepackage{bm}
\usepackage{adjustbox}
\usepackage{multirow}
\usepackage{mathrsfs}
\usepackage{adjustbox}
\usepackage{amssymb}
\usepackage{slashed}    
\usepackage{caption}     
\usepackage{subcaption}    
\usepackage{wrapfig} 
\usepackage[normalem]{ulem}

\def\beq{\begin{equation}}
\def\eeq{\end{equation}}
\def\bea{\begin{eqnarray}}
\def\eea{\end{eqnarray}}

\usepackage{array}
\newcolumntype{P}[1]{>{\centering\arraybackslash}p{#1}}
\newcolumntype{M}[1]{>{\centering\arraybackslash}m{#1}}

\definecolor{lightblue}{rgb}{0.1, 0.5, 1.0}
\definecolor{darkblue}{cmyk}{1,0.4,0,0.3}
\definecolor{violet}{cmyk}{0,1,0,0.2}

\newcommand{\blue}[1]{\color{blue}#1\color{black}}

\RequirePackage[
colorlinks=true,
urlcolor=blue,
anchorcolor=blue,
citecolor=blue,
filecolor=blue, 
linkcolor=blue,
menucolor=blue,
linktocpage=true,
pdfproducer=medialab,
pagebackref=true
]{hyperref}
\hypersetup{
  colorlinks   = true, 
  urlcolor     = mygreen, 
  linkcolor    = blue, 
  citecolor   = blue 
}
\usepackage{orcidlink}
\usepackage[font=small,labelfont=bf,justification=centerlast,singlelinecheck=true]{caption}
\usepackage{mathtools}

\usepackage[capitalise]{cleveref}
\crefname{section}{Sec.}{Sec.}

\usepackage{csquotes}

\definecolor{mygreen}{rgb}{0.0, 0.5, 0.0}
\definecolor{blus}{cmyk}{1,1,0,0.6}

\newcommand{\bld}[1]{\boldsymbol{#1}}
\newcommand{\tbf}[1]{\textbf{#1}}
\newcommand{\ovbld}[1]{\overline{\boldsymbol{#1}}}
\newcommand{\nn}{\nonumber}
\newcommand{\op}[1]{\operatorname{#1}}

\newcommand{\LL}{\mathcal{L}}

\newcommand{\OO}{\mathcal{O}}

\begin{document}

{\flushright
{\blue{ \hfill}\\
\blue{ \hfill}\\
\blue{MIT-CTP/5768}\\
\blue{IFT-UAM/CSIC-25-10}\\
}}

\begin{center}
{\LARGE\bf\color{blus} 
Higgs near-criticality at future colliders\\
}
\vspace{1cm}

{\bf 
Victor Enguita\orcidlink{0000-0001-5977-9635}$^{a}$, 
Belen Gavela\orcidlink{0000-0002-2321-9190}$^{a}$, 
Thomas Steingasser\orcidlink{0000-0002-1726-2117}$^{b,c}$}\\[7mm]

{\it $^a$Departamento de Fisica Teorica, Universidad Autonoma de Madrid, \\ 
and IFT-UAM/CSIC, Cantoblanco, 28049, Madrid, Spain}\\[1mm]
{\it $^b$Department of Physics, Massachusetts Institute of Technology, Cambridge, MA 02139, USA}\\[1mm]
{{\it $^c$Black Hole Initiative at Harvard University, 20 Garden Street, Cambridge, MA 02138, USA}}

\vspace{0.5cm}

\begin{abstract}

The so-called metastability bound on the Higgs mass suggests that the smallness of the Higgs mass may be a byproduct of the metastability of the electroweak vacuum.  
A significantly strong bound requires new physics capable of lowering the scale where the Higgs quartic coupling turns negative through renormalization group effects, without destabilizing the electroweak vacuum entirely. We analyze in this context the low-scale Majoron model of neutrino masses, which automatically contains two key elements for a viable scenario: heavy fermions to lower the instability scale and a extended scalar sector to stabilize the potential and achieve realistic lifetimes for the electroweak vacuum. We show how the metastability bound can be generalized to theories with multiple scalars and present an efficient way of calculating the tunneling rate in such potentials. We also demonstrate that FCC will probe regions of the parameter space relevant for metastability: large regions of the fermionic sector at FCC-ee and some reach to the scalar sector at FCC-hh.

\end{abstract}

\vspace{1cm}
{\emph{Email:} \url{victor.enguita@uam.es}, \url{belen.gavela@uam.es}, \linebreak \url{tstngssr@mit.edu}}

\thispagestyle{empty}
\bigskip

\end{center}

\setcounter{footnote}{0}

\newpage
\tableofcontents

\pagebreak

\section{Introduction}
\label{sec:intro}

The present Higgs quest is largely motivated by the so-called \textit{naturalness} problem: the fact that the value of the Higgs mass is comparable to the electroweak (EW) scale while quantum effects generically try to push it to high values, {\bf if} there are new physics scales beyond the EW one to which the Higgs mass is sensitive. Within the naturalness paradigm, small and/or apparently fine-tuned parameters would be explained away as being required by underlying symmetries. A plethora of beyond-the-Standard-Model (BSM) scenarios trying to justify the smallness of the Higgs mass through new symmetries has been intensely studied since decades, with no real breakthrough by now and no experimental evidence for the associated scales of new physics. A similar puzzle can be found when analyzing the stability of the Higgs field. Its relevant parameters can be interpreted as fine-tuned to allow this field to decay into an energetically more favorable state, destroying the Universe as we know it in the process, but only with an exceedingly small probability. Altogether, the paradigm of naturalness together with the properties of the Higgs boson would suggest that the LHC should have already found signs of new physics beyond the Standard Model (BSM), to justify those values through symmetry arguments.

The situation is so perplexing that it has  been questioned whether the principle of naturalness  should be applied to the Higgs system at all.
An alternative paradigm is that of {\it self-organized criticality}~\cite{bak1987self} and related developments,  which would drive the parameters of the scalar potential of Nature  towards {\it near-criticality}, i.e., towards  apparently fine-tuned values.  
Indeed, the apparent fine-tuned values of the two independent parameters of the SM Higgs potential can be understood as close to critical values, i.e., the Higgs mass is very small (the limit $m_h \rightarrow 0$ separates the phases with and without spontaneous symmetry breaking) and the value of the Higgs self-coupling  $\lambda$,  i.e., the strength with which the Higgs field scatters off itself, is tantalizingly close to the transition from a stable to an unstable one, as mentioned above.
 To require those two parameters to be close to critical 
  amounts to a drastically unusual perspective on the hierarchy problem: rather than directly trying to explain the relative lightness of the Higgs, one might instead try to focus on the mechanism responsible for favoring metastability, and find a small Higgs mass as a byproduct.  
 For concrete examples of physical underlying mechanisms  see Refs.~\cite{Giudice:2021viw,Khoury:2019ajl,Khoury:2019yoo,Kartvelishvili:2020thd,Khoury:2021grg},  which speculate that the near-criticality of the Higgs potential parameters could be the result of a dynamical mechanism in the early Universe.\footnote{
Beyond the vacuum selection mechanism developed in these references,  it was recently argued in Ref.~\cite{Benevedes:2024tdq} that a metastability bound also emerges naturally in any theory whose vacuum manifold (or landscape) satisfies certain generic conditions. }

This work will remain agnostic and it will not rely on any particular mechanism favoring near-criticality,  but it will rather build upon the interdependence of the parameters of the scalar potential required by near-criticality. Amongst the previous efforts in this direction  are the \textit{metastability bounds} for the Higgs system developed in Refs.~\cite{Buttazzo:2013uya,Khoury:2021zao,Benevedes:2024tdq}, and this is the path adopted in this work. The metastability bounds represent a significant deviation from usual model-building efforts: rather than trying to construct an explicit model capable of explaining the apparent fine-tunings of the potential, these bounds establish an interdependence between them. More concretely, they state that the metastability of the EW arising from the renormalisation group (RG) flow of the quartic coupling $\lambda$ to negative values at high energies  \textit{imposes} an upper limit on the Higgs mass. This suggests that finding an explanation for the apparent metastability could also offer a solution to the EW hierarchy problem.

Crucially, the metastability bounds do not depend on the details of the mechanism responsible for the vacuum's metastability. This insensitivity establishes the universality of the bounds and makes transparent their most important dependencies.  

The precise value of the metastability bound on the Higgs mass is roughly set by the instability scale $\mu_I$,  defined as the scale at which $\lambda$ vanishes and then becomes negative when loop corrections are taken into account. In the SM -- see; for instance, Refs.~\cite{Buttazzo:2013uya,Khoury:2021zao,Chauhan:2023pur,Steingasser:2023ugv}, 
\begin{align}
    m_h^2 <  |\beta_{\lambda}| e^{-3/2} \cdot \mu_I^2 
    \,,
    \label{eq:metastability-bound}
\end{align}
where $\beta_\lambda$ is the beta function for $\lambda$. Intuitively, Eq.~(\ref{eq:metastability-bound}) suggests that the upper limit on $m_h$ is  about one order of magnitude smaller than $\mu_I$.   In the SM $\mu_I \sim 10^{11}$ GeV, which translates into a SM bound $m_h^2  \leq 10^{10}$ GeV, which is a very weak and in practice irrelevant constraint.
  
Nevertheless, the bound may be very sensitive  to  BSM matter contents, which can strongly lower the value of $\mu_I$ down to values $\mathcal{O}(1)$~TeV, allowed by current colliders.   
Indeed, it has been shown that strict bounds on $m_h$ can emerge in the presence of heavy exotic fermions, which strongly affect the renormalization group (RG) running of $\lambda$ and lower the instability scale. However, a byproduct of this mechanism to lower the instability scale is a further destabilization of the vacuum: for couplings strong enough to get the bound close to the observed value of the Higgs mass, the effect is also strong enough to lower the vacuum's lifetime $\tau$ below the current age of the Universe, contradicting our ongoing existence. This establishes another necessary condition for the metastability bound's ability to explain the Higgs mass, namely the need for some additional new physics capable of partially stabilizing the vacuum at energies above the instability scale. In summary, viable scenarios that achieve a strict metastability limit on the Higgs mass can be achieved with the combination of:
\begin{itemize}
\item Heavy fermions to lower the instability scale $\mu_I$ so that the limit on $m_h$ is close to the observed value.
\item At least one BSM scalar to stabilize the vacuum at a scale $\mu_S > \mu_I$, so as to raise the vacuum’s lifetime above the current age of the Universe.
\end{itemize}
Note that this very specific combination and ordering of new physics scales is, as of now, only favored by the metastability bound. Thus, the discovery of such a pattern would provide circumstantial evidence for its importance in explaining the observed value of the Higgs mass.

An excellent example of suitable fermions are heavy right-handed neutrinos, nowadays often dubbed as heavy neutral leptons (HNLs), such as those present in seesaw models of neutrino masses~\cite{Minkowski:1977sc,Yanagida:1979as,Glashow:1979nm,Mohapatra:1979ia,Magg:1980ut,Schechter:1980gr,Wetterich:1981bx,Lazarides:1980nt,Mohapatra:1980yp,Ma:1998dn}.~\footnote{Non-singlet exotic fermions have also been considered, for instance in Ref.~\cite{Steingasser:2023ugv}.} The destabilizing effect of exotic heavy fermions via their lowering of $\mu_I$ has been first pointed out in Ref.~\cite{Casas:1999cd}. The stabilizing counter-effect of scalars was first studied in Ref.~\cite{Elias-Miro:2012eoi}. Many other works have studied both effects (see for instance \cite{DelleRose:2015bms,Lindner:2015qva,Bambhaniya:2016rbb,Lebedev:2012zw}), including in the Majoron model~\cite{Chikashige:1980ui} setup (see, e.g., Refs.~\cite{Sirkka:1994np,Bonilla:2015kna, Mandal:2019ndp}), although with the aim of stabilizing the SM framework and without connection to a possible Higgs mass limit. 

In the context of the Higgs mass metastability bound, type-I seesaw fermions were considered in Refs.~\cite{Steingasser:2023ugv, Khoury:2021zao,Benevedes:2024tdq}, while the effect of scalar physics  has been explored via a mass-dimension six ($d=6$) scalar operator~\cite{Steingasser:2023ugv}, in the spirit of an Effective Field Theory (EFT). At leading order, the impact of BSM physics appears then via two parameters: the value of $\mu_I$ and the scale -- larger than the EW one -- suppressing the $d=6$ correction to the Higgs' potential (although, in practice, Eq.~(\ref{eq:metastability-bound}) still provides then the dominant contribution to the bound, see below). This EFT approach has the advantage of model-independence at the expense of predictability. It renders difficult the interpretation of the results in terms of multiple scalars, and it also fails to incorporate the effect of other $d=6$ operators such as momentum-dependent corrections. More importantly, the leading scalar effective operators may not suffice in the regime where the $m_h$ bound is strongest, as the lifetime calculation becomes then very dependent on the tower of higher-dimensional operators, thus making the tunneling rate calculation sensitive to the full structure of the scalar potential. To address this point we explore here the metastability  bound on $m_h$ in an ultraviolet complete model: the Majoron completion of the type-I seesaw mechanism.

The Majoron completion~\cite{Chikashige:1980ui} is particularly suited for our study, as it naturally contains both ingredients in the bullet points above: HNLs plus a scalar field whose vacuum expectation value (vev) sets the overall scale of the heavy Majorana neutrino masses $m_N$ and also of the scalar mass $M_s$. This fact automatically addresses the subtle question of the proximity of $\mu_I$ and the new physics scale to obtain a tight bound. To be precise, it encompasses the condition
\begin{equation}
m_N < \mu_I < M_s\,,
\label{scales}
\end{equation}
with all these scales naturally close within a few orders of magnitude because of their common origin. We will also show that in this model metastability genuinely \textit{requires} the quartic coupling to be negative at the RG scale corresponding to the scalar mass, and then run towards a positive value at lower energies. 

In order to obtain strict limits on $m_h$,  
a Majoron model with a scale comparable to or only a few orders of magnitude above $m_h$ is thus pertinent.  This immediately raises the question of how to obtain realistically light neutrino masses. The solution lies within the large class of so-called ``low-scale'' Majoron models, which exhibit an approximate $U(1)$ lepton-number symmetry: the symmetry suppresses the size of light neutrino masses even with BSM scales near the EW one~\cite{Branco:1988ex,Kersten:2007vk,Abada:2007ux,Moffat:2017feq, Mohapatra:1986aw,Mohapatra:1986bd,Akhmedov:1995ip,Malinsky:2005bi,Shaposhnikov:2006nn}. 
In other words, it decouples the size of the light neutrino masses from the rest of the analysis. 

Finally, an important question on any serious alternative to the naturalness paradigm is its testability at proposed future accelerators. This requires to tackle the global shape of the Higgs potential, i.e., to go beyond its second derivative (the Higgs mass), which only determines the local curvature and is already known with better than percent accuracy. Indeed, future colliders aim to measure the third derivative of the potential (i.e., the Higgs self-coupling) with  percent level accuracy, so as to start to resolve and probe the quantum structure of the potential.  
We will analyze the  signals at the FCC-ee and the FCC-hh  expected from the Majoron model, in the parameter range that leads to  strict metastability bounds on $m_h$ and viable Universe lifetimes. 

The data from future colliders offers an excellent window of opportunity,  as the reference scale $\mu_I$ in Eq.~(\ref{scales}) close to $m_h$ suggests the need to consider dynamical heavy neutrinos at reach of the FCC-ee, for whose energy range the exotic scalar sector can be integrated out and treated via effective field theory (EFT). When facing the FCC-ee prospects we will thus profit from previous studies on:  i) limits on the mixing of heavy and light neutrinos; ii)  one-loop analyses of the Majoron model and its impact on the values of SMEFT operator coefficients once the exotic scalar sector is integrated out~\cite{Jiang:2018pbd}; iii) very precise sensitivity studies of the FCC-ee to the SMEFT operator coefficients~\cite{Celada:2024mcf}. For the laboratory data on the energy range of interest for the scalar sector, we will also use EFT techniques, 
relying  on previous numerical studies of the FCC-hh sensitivity to $\lambda$~\cite{Mangano:2020sao}.

For the RG analyses above the exotic scalar mass range, the complete Majoron model structure must be considered instead. The same applies to the vacuum lifetime analyses.

\section{The low scale Majoron model}
\label{sec:majoron-completion}

In the \emph{Majoron} model~\cite{Chikashige:1980ui}, 
light neutrino masses result from a Type-I seesaw scenario in which lepton number $L$ is spontaneously violated via the vacuum expectation value (VEV) of an additional complex scalar field $S = |S| e^{i\,\phi/\Lambda}$, where $S$ denotes the radial component and $\phi$ the (pseudo-)Goldstone boson -- the Majoron. Given Eq.~(\ref{scales}), two distinct regimes of energies $E$ are of particular interest:
\begin{itemize}
    \item{$E>M_s$, for which the completed ultraviolet model needs to be considered.}
    \item{$M_N<E<M_s$}, in which the heavy neutrinos $N$ are dynamical, while the $S$ field can be integrated out. This regime can be appropriately described by the combination of the  SMEFT Lagrangian (i.e., SM Lagrangian plus effective operators made out of SM fields), the type-I seesaw Lagrangian and additional operators made out of $N$ and the SM fields. The total Lagrangian is sometimes referred to as the \textit{$\nu$SMEFT Lagragian}.
\end{itemize}
We detail next the Lagrangians for these two regimes.

\subsection{Scalar sector}
\label{sec:scalar}
For energies $E>M_s$, the  relevant scalar Lagrangian reads
\begin{align}
    &\mathcal{L}= (D_\mu H)^\dagger D^\mu H + (D_\mu S)^\dagger D^\mu S - V(H,S)\,,
\end{align}
with 
\begin{align}
    &V(H, S)=-\mu_H^2 |H|^2 -\mu_S^2 |S|^2+
\lambda_H |H|^4 +\lambda_S |S|^4+\kappa\,  |H|^2 |S|^2\,+\, V_{\slashed{L}}   \,,
\label{eq:majoron:potential}
\end{align}
where $H$ and $S$ denote respectively the SM Higgs doublet and the complex scalar singlet, with  $|H|^2\equiv H^\dagger H$ and $|S|^2\equiv S^* S$, and  $ V_{\slashed{L}}$  denotes some small and explicitly L-breaking term to render the pseudo-Goldstone Majoron massive (we will disregard this term  from now on, as its specific form is irrelevant for our study). Upon spontaneous breaking of the electroweak symmetry (EWSB) and of lepton number symmetry, $S$ and $H$ take VEVs to be denoted respectively by $w$ and $v$, 
\begin{equation}
|S| \rightarrow \frac{w + s}{\sqrt{2}}\,,\qquad |H| \rightarrow \frac{v + h}{\sqrt{2}}\,,
\label{vevs}
\end{equation}
where the Higgs doublet has been expressed in unitary gauge and $s$ and $h$ denote the physical scalar excitations.

We will study below the conditions required for this potential to lead to a metastability bound for the Higgs mass. To this aim, its quantum phases will be classified by the number of vacua and their properties, as the emergence/disappearance of a minimum is key to that bound. Indeed, a necessary condition for metastability is the development of at least two minima: a metastable vacuum at the electroweak scale and a true vacuum at some higher energy. We will thus first analyze the tree-level stationary points of the potential, paying attention to its saddle points: they can seed the metastable EW vacuum once radiative corrections will be taken into account in a second step.

\begin{figure}
\centering
\begin{subfigure}{0.4\textwidth}
\centering
{$\hspace{0.8cm}\bld{\mu_H^2, \mu_S^2 >0}$}\\
\includegraphics[width = \textwidth]{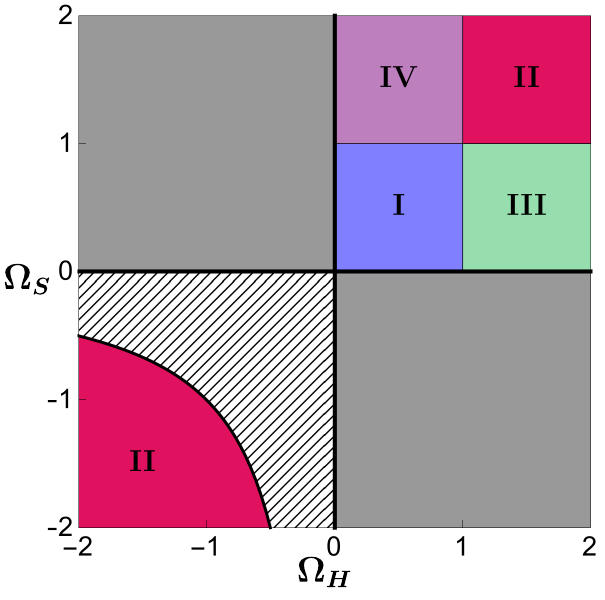}
\end{subfigure}
\begin{subfigure}{0.4\textwidth}
\centering
{ $\hspace{0.7cm}\bld{\mu_H^2 < 0  < \mu_S^2}$}\\
\includegraphics[width = \textwidth]{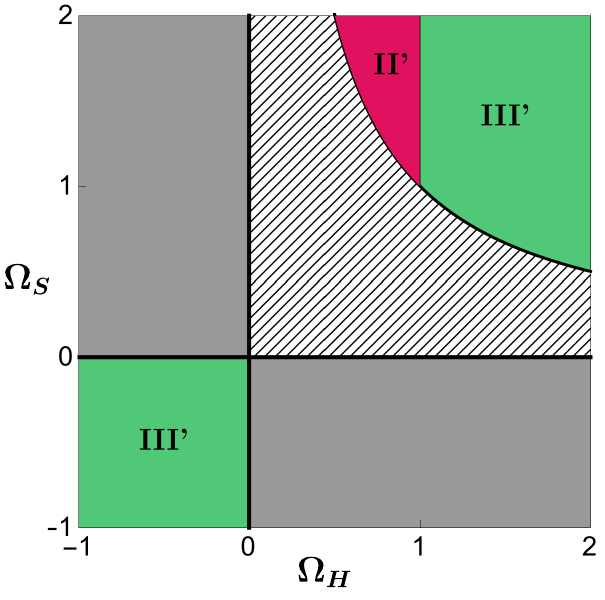}
\end{subfigure}
\begin{subfigure}{0.4\textwidth}
\centering
{$\hspace{0.7cm}\bld{\mu_H^2 > 0  > \mu_S^2}$}\\
\includegraphics[width = \textwidth]{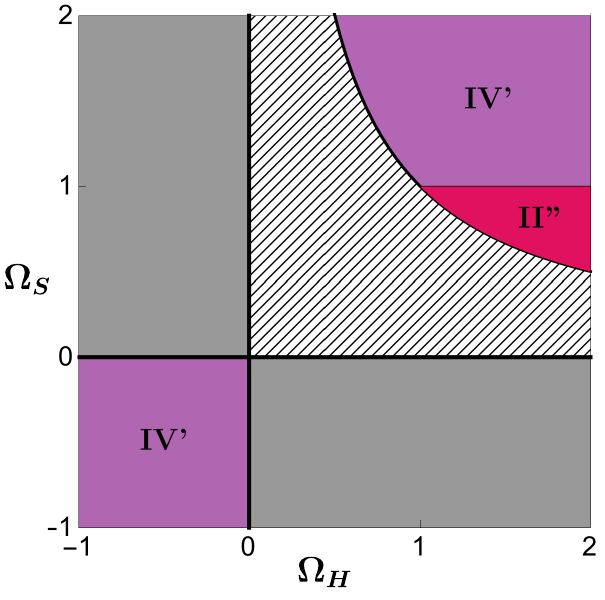}
\end{subfigure}
\begin{subfigure}{0.4\textwidth}
\centering
{ $\hspace{0.8cm}\bld{\mu_H^2, \mu_S^2 < 0}$}\\
\includegraphics[width = \textwidth]{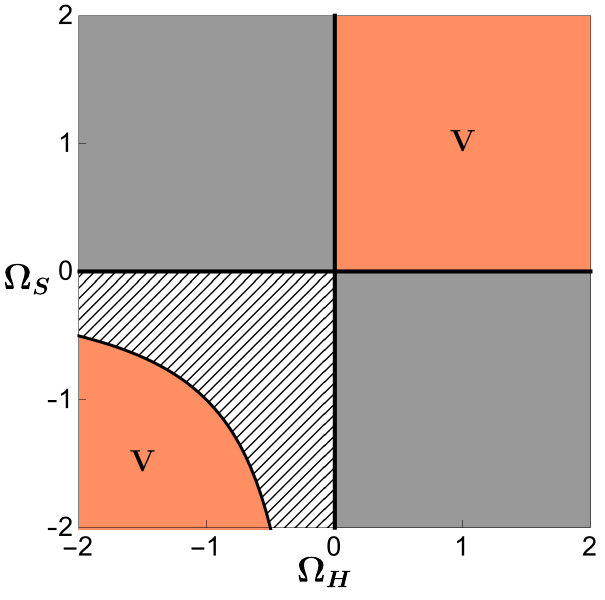}
\end{subfigure}
\caption{
\tbf{Phase space $\bld{(\Omega_H, \Omega_S)}$}. The Phases \textbf{I}-\textbf{V}, \tbf{II'}-\tbf{IV'} and \tbf{II''} correspond to all possible configurations of the stationary points of the potential in \cref{eq:majoron:potential} under the assumption of stability. Grey regions are not accessible under that assumption. Hatched regions are unstable, see text.}
\label{fig:potential:phase-space-diagrams}
\end{figure}

\begin{figure}[hbpt!]
\begin{subfigure}{0.32\textwidth}
\centering
\hspace*{0.5cm} \textbf{\footnotesize (a) Phase I}\vspace{0.05cm}
\includegraphics[width = \textwidth]{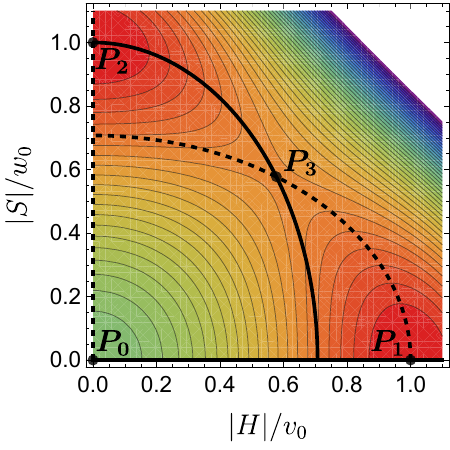}{}
\label{fig:potential:configurations:PhaseI}
\end{subfigure}
\begin{subfigure}{0.32\textwidth}
\centering
\hspace*{0.5cm} \textbf{\footnotesize (b) Phase II}\vspace{0.05cm}
\includegraphics[width = \textwidth]{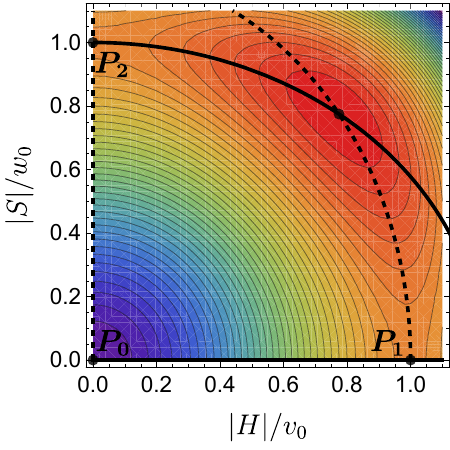}{}
\label{fig:potential:configurations:PhaseII}
\end{subfigure}
\begin{subfigure}{0.32\textwidth}
\centering
\hspace*{0.5cm} \textbf{\footnotesize (c) Phase II'}\vspace{0.05cm}
\includegraphics[width = \textwidth]{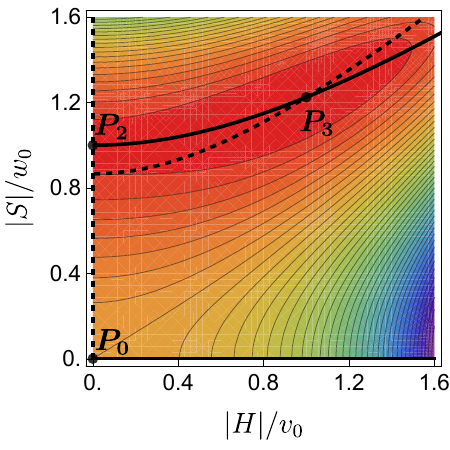}{}
\label{fig:potential:configurations:PhaseIIprime}
\end{subfigure}
\begin{subfigure}{0.32\textwidth}
\centering
\hspace*{0.5cm} \textbf{\footnotesize (d) Phase II''}\vspace{0.05cm}
\includegraphics[width = \textwidth]{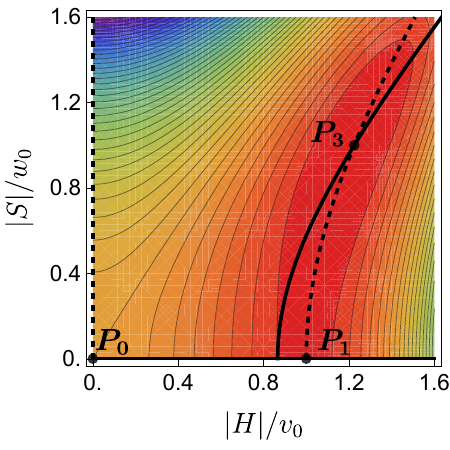}{}
\label{fig:potential:configurations:PhaseIIprime2}
\end{subfigure}
\begin{subfigure}{0.32\textwidth}
\centering
\hspace*{0.5cm} \textbf{\footnotesize (e) Phase III}\vspace{0.05cm}
\includegraphics[width = \textwidth]{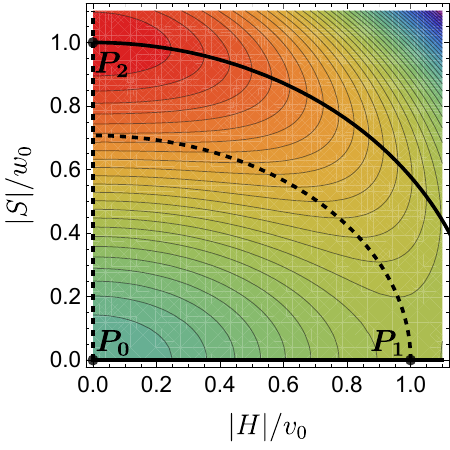}{}
\label{fig:potential:configurations:PhaseIII}
\end{subfigure}
\begin{subfigure}{0.32\textwidth}
\centering
\hspace*{0.5cm} \textbf{\footnotesize (f) Phase III'}\vspace{0.05cm}
\includegraphics[width = \textwidth]{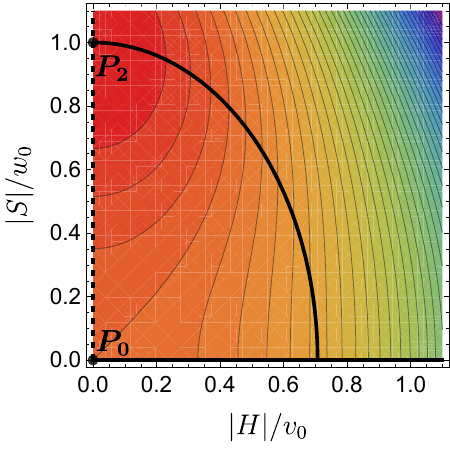}{}
\label{fig:potential:configurations:PhaseIIIp}
\end{subfigure}
\begin{subfigure}{0.32\textwidth}
\centering
\hspace*{0.5cm} \textbf{\footnotesize (g) Phase IV}\vspace{0.05cm}
\includegraphics[width = \textwidth]{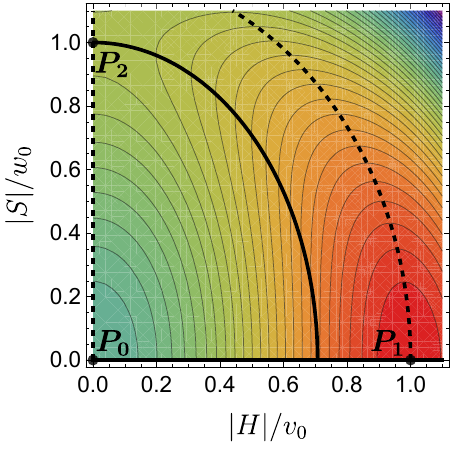}{}
\label{fig:potential:configurations:PhaseIV}
\end{subfigure}
\begin{subfigure}{0.32\textwidth}
\centering
\hspace*{0.5cm} \textbf{\footnotesize (h) Phase IV'}\vspace{0.05cm}
\includegraphics[width = \textwidth]{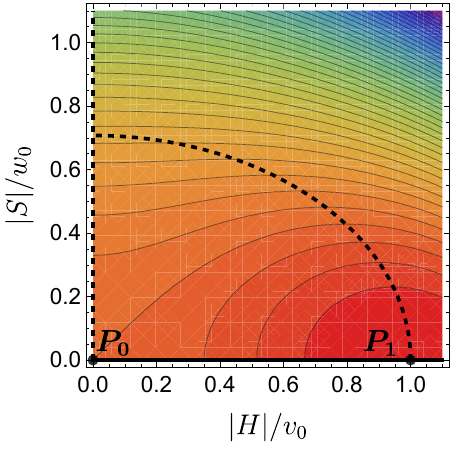}{}
\label{fig:potential:configurations:PhaseIVprime}
\end{subfigure}
\begin{subfigure}{0.32\textwidth}
\centering
\hspace*{0.5cm} \textbf{\footnotesize (i) Phase V}\vspace{0.05cm}
\includegraphics[width = \textwidth]{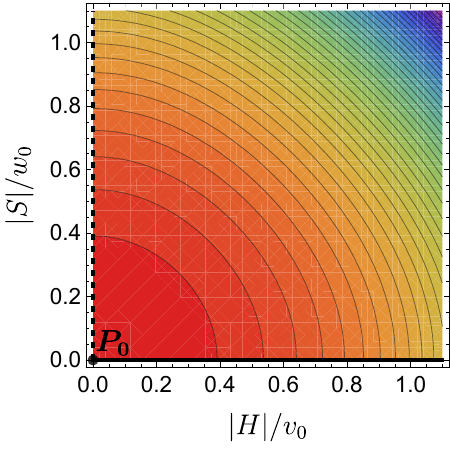}{}
\label{fig:potential:configurations:PhaseIIIprime}
\end{subfigure}
\caption{Examples of configurations belonging to phases \tbf{I}-\tbf{V} and \tbf{II'}-\tbf{IV'} and \tbf{II''}. 
The black solid and dashed lines correspond respectively  to the paths defined by the two  ellipses in Eqs.~(\ref{Hminima}) and (\ref{Sminima}) 
as well as the axes $H = 0$ and $S = 0$.}
\vspace*{-0.5 cm}
\tiny
\begin{align*}
\raggedleft
&\text{\scriptsize\emph{Parameters used:}} \\
&\begin{array}[t]{cc}
\begin{array}[t]{rcccccc}
           & (\Omega_{H}, \Omega_{S})
           & \mu_{H}^2{\,\tiny \text{ (GeV}^{2})}
           & \mu_{S}^2{\,\tiny \text{(GeV}^{2})}
           & \lambda_H
           & \lambda_S
           & \kappa \\
  \tbf{I:} & (1/2,1/2) 
           & \,\,\,\,1
           & \,\,\,\,1
           & 1/4
           & 1/4
           &  \,\,\,\,1 \\
 \tbf{II:} & (3/2,3/2) 
           & \,\,\,\,1 
           & \,\,\,\,1 
           & 3/4
           & 3/4 
           & \,\,\,\,1 \\
\tbf{II':} & (2,3/4) 
           & -1
           & \,\,\,\,1
           & 1
           & 3/8 
           & -1 \\
\tbf{II'':} & (3/4,2) 
           & \,\,\,\,1
           & -1
           & 3/8
           & 1
           & -1 \\
\tbf{III:} & (3/2,1/2) 
           & \,\,\,\,1
           & \,\,\,\,1
           & 3/4
           & 1/4
           &  \,\,\,\,1 \\
\end{array}
&
\begin{array}[t]{rcccccc}
           & (\Omega_{H}, \Omega_{S})
           & \mu_{H}^2{\,\tiny \text{ (GeV}^{2})}
           & \mu_{S}^2{\,\tiny \text{(GeV}^{2})}
           & \lambda_H
           & \lambda_S
           & \kappa \\
\tbf{III':} & (-1/2, -1/2) 
           & -1
           & \,\,\,\,1
           & 1/4
           & 1/4
           & 1 \\
\tbf{IV:} & (1/2,3/2) 
           & \,\,\,\,1
           & \,\,\,\,1
           & 1/4
           & 3/4
           & 1\\
\tbf{IV':}& (-1/2, -1/2)
           & \,\,\,\,1 
           & -1
           & 1/4
           & 1/4 
           & 1 \\
\tbf{V:} & (1/2, 1/2) 
           & -1 
           & -1
           & 1/4
           & 1/4 
           & 1 \\
\end{array}
\end{array}
\end{align*}
\label{fig:potential:configurations}
\end{figure}

For the potential in \cref{eq:majoron:potential} and at tree-level, the stationary points $\bld{P}_k \equiv (H_k,\,S_k)$ are solutions of
\begin{align}
0 = \frac{\partial V}{\partial |H|} &= 2\,|H|\cdot \left( - \mu_H^2 + \kappa |S|^2 + 2\,\lambda_H |H|^2 \right)\,,\\
0= \frac{\partial V}{\partial |S|} &= 2\,|S|\cdot \left( - \mu_S^2 + \kappa |H|^2 + 2\,\lambda_S |S|^2 \right)\,.\label{eq:<S>}
\end{align}
The possible minima thus correspond to combinations of the two following sets of solutions: 
\begin{align}
	&\partial_{|H|} V=0:  
        &&\Bigg\{ \quad
        |H|=0 \quad
        , \quad
        2\,\lambda_H\,\frac{|H|^2}{\mu_H^2} + \,\kappa\,\frac{|S|^2}{\mu_H^2} = 1\quad
        \Bigg\}\,, \label{Hminima}\\
	&\partial_{|S|} V=0:
        &&\Bigg\{  \quad 
        |S|=0 \quad
        , \quad \,\,\,2\,\lambda_S\,\frac{|S|^2}{\mu_S^2}+\kappa\,\frac{|H|^2}{\mu_S^2} = 1 \quad 
        \Bigg\}\,,
        \label{Sminima}
\end{align}  
where in these solutions $|H|$ and $|S|$ stand for their respective {VEV}s. These equations lead to a set of four stationary points $\bld{P}_k$,
\begin{align}
    &\bld{P}_0 \equiv (0,\,0   )\,, 
    &&\bld{P}_1 \equiv \left(\mu_H/(2\,\lambda_H)^{1/2},\,0   \right)\,, 
    &&\bld{P}_2 \equiv \left(0,\,\mu_S/(2\,\lambda_S)^{1/2}\right)\,,
    \label{P0-P1-P2}
\end{align}
\begin{align}
    &\bld{P}_3  = 
    \left( \sqrt{
    \frac{2\,\mu_H^2\,\lambda_S - \mu_S^2\,\kappa}{4\,\lambda_H\,\lambda_S - \kappa^2}}
    \,,
 \sqrt{\frac{2\,\mu_S^2\,\lambda_H - \mu_H^2\,\kappa}{4\,\lambda_H\,\lambda_S - \kappa^2}}
    \right) \,.
    \label{P3}
\end{align} 
Whenever they correspond to minima, their locations correspond to the VEVs $v$ and $w$ in Eq.~(\ref{vevs}). A simple geometric interpretation of their location has been given in Ref.~\cite{Espinosa:2011ax}: in addition to the origin of coordinates in the $\{|H|,|S|\}$ plane, the $\bld{P}_k$'s lie at the intersections of the two ellipses described by the non-trivial solutions in Eqs.~(\ref{Hminima}) and (\ref{Sminima}), and by their intersection with the $|S|$ and $|H|$ axes, respectively. It is interesting to recast their expressions in terms of normalized stationary points:  
\begin{align}
&\ovbld{P}_k = \sqrt{2}\,(|H|_k/v_0,\, |S|_k/w_0) 
&&\text{with} 
&&v_0^2\equiv |\mu_H^2|/\lambda_H
\,,
&&w_0^2\equiv |\mu_S^2|/ \lambda_S\,,
\end{align}
where, in the limit $\kappa \to 0$,  $v_0$ and $w_0$ would respectively be the only non-trivial VEVs of the two scalar fields $H$ and $S$, and where the normalization in \cref{vevs} has been used. The tree-level stationary points of the complete theory then read
\begin{align}
    &\ovbld{P}_0 = \left(0, 0\right)\,,
    &&\ovbld{P}_1 = \left(\sqrt{\op{sign}[\mu_H^2]},\,0\right)\,,
    &&\ovbld{P}_2 = \left(0,\,\sqrt{\op{sign}[\mu_S^2]}\right)\,,
    \label{eq:bar-P0-P1-P2}
\end{align}
\begin{align}
    &\ovbld{P}_3 = 
    \left(
    \sqrt{\op{sign}[\mu_H^2]\,\frac{\Omega_H\,(\Omega_S-1)}{\Omega_H\,\Omega_S-1}}
    ,\,
    \sqrt{\op{sign}[\mu_S^2]\,\frac{\Omega_S\,(\Omega_H-1)}{\Omega_H\,\Omega_S-1}}
    \right)\,,
     \label{eq:bar-P3}
\end{align}
where 
\begin{align}
    &\Omega_H \equiv 2\,\frac{\mu_S^2}{\mu_H^2} \cdot \frac{\lambda_H}{\kappa}\,,\ &&\text{and}   
    &&\Omega_S \equiv 2\,\frac{\mu_H^2}{\mu_S^2} \cdot \frac{\lambda_S}{\kappa}\,.
    \label{eq:potential:definition-Omega_H-Omega_S}
\end{align}
It follows that the nature of the extrema (whether minima, maxima or saddle points) is
\emph{only} controlled by $\Omega_H$, $\Omega_S$ and the signs of $\mu_H^2$ and $\mu_S^2$. Demanding the potential to be bounded from below further requires 
\begin{align} 
     \lambda_H, \lambda_S > 0\,,\qquad &\text{ for} \qquad \kappa\ge 0\,, \label{eq:potential:global-stability-conditions}\\
     \lambda_H, \lambda_S > 0\ \text{ and }\ \lambda_S\,\lambda_H >\frac{\kappa^2}{4}\,
     ,\qquad &\text{ for} \qquad \kappa < 0\,.
     \label{eq:potential:global-stability-conditions-negative}
 \end{align}
All possible contours in the $\{|H|,|S|\}$-plane are depicted in \cref{fig:potential:phase-space-diagrams} for any combination of  signs of the parameters: phases \tbf{I}-\tbf{V}, \tbf{II'}-\tbf{IV'} and \tbf{II''} constitute \emph{all} possible bounded-from-below  configurations of the potential, defined through qualitatively different physical behaviors. Extrema in each of the phases are classified in \cref{tab:potential:extrema-classification} according to their nature  as maxima, minima or saddle points -- see \cref{sec:appendix:phase-space-2-field-potential} for further details --  while the corresponding potentials are depicted in \cref{fig:potential:configurations}. The ``primed'' notation points to phases which can be seen as ``variations'' of unprimed ones, as they share the same minima while one of the saddle points changes. The boundness conditions in Eqs.~(\cref{eq:potential:global-stability-conditions})-(\ref{eq:potential:global-stability-conditions-negative}) eliminate a part of the phase space: the first one implies that $\Omega_H$ and $\Omega_S$ have the same sign so that the $(+,-)$ and $(-, +)$ quadrants -- depicted in grey in \cref{fig:potential:phase-space-diagrams} -- are not accessible. Some quadrants can only be accessed by setting $\kappa < 0$ and the black and white hatched regions represent the sector within those regions that is forbidden by  \cref{eq:potential:global-stability-conditions-negative}. Phase \tbf{V} is of no interest to us as it exhibits no spontaneous symmetry breaking. 

The parameters $|\Omega_H|$ and $|\Omega_S|$ also have a geometric interpretation in terms of the ellipses in Eqs.~(\ref{Hminima}) and (\ref{Sminima}) as the ratio of the square of the distance to the origin of their intersection points with the $|H|$ and $|S|$ axis, respectively. Assume for illustration any case in which all parameters in the potential are positive. If then, for example, $\Omega_H,\Omega_S>1$ or $\Omega_H,\Omega_S<1$, the contours representing solutions to Eqs.~(\ref{Hminima})-(\ref{Sminima}) must intersect in the bulk of the $(|H|,|S|)$-plane (i.e., away from the axes), leading to the formation of a non-trivial local extremum in this plane, as in contours \tbf{I} and \tbf{II} in \cref{fig:potential:configurations}. If, on the other hand, only $\Omega_H > 1$ or $\Omega_S > 1$, these contours only intersect with the axes, e.g. \tbf{III} and \tbf{IV} in that figure. An analogous analysis applies to other signs of the potential parameters. 

Such a tree-level picture of the landscape of possible configurations in a two-scalar potential may change when considering one-loop effects, i.e., under RG evolution (RGE). In particular, a potential barrier may develop around some of the saddle points shown above, transforming them into extra local minima. This possibility is exploited in the next section to realize an interesting scenario showcasing EW vacuum metastability.
\begin{table}
\centering
\begin{align*}
\begin{array}{c|c|c|c|c|c|c|c|c|c}
&  \text{\tbf{I}} 
& \text{\tbf{II}} 
& \text{\tbf{II'}}
& \text{\tbf{II''}}
& \text{\tbf{III}}
& \text{\tbf{III'}} 
& \text{\tbf{IV}}
& \text{\tbf{IV'}}
& \text{\tbf{V}} \\
\hline
\text{\textsc{maxima}}
&  \bld{P_0}
&  \bld{P_0}
&  -
&  -
&  \bld{P_0}
&  -
&  \bld{P_0}
&  -
&  -\\
\hline
\text{\textsc{minima}}
&  \bld{P_1},\bld{P_2}
&  \bld{P_3}
&  \bld{P_3}
&  \bld{P_3}
&  \bld{P_2}
&  \bld{P_2}
&  \bld{P_1}
&  \bld{P_1}
&  \bld{P_0}\\
\hline
\text{\textsc{saddles}}
&  \bld{P_3}
&  \bld{P_1}, \bld{P_2}
&  \bld{P_0}, \bld{P_2}
&  \bld{P_0}, \bld{P_1}
&  \bld{P_1}
&  \bld{P_0}
&  \bld{P_2}
&  \bld{P_0} 
&  -\\
\end{array}
\end{align*}
\caption{Configuration on the extrema in all the possible phases under the assumption of boundedness from below.}
\label{tab:potential:extrema-classification}
\end{table}

\subsection{Neutrino masses}
It has been shown that heavy right-handed neutrinos $\nu_{R_i}$ can be crucial to lower the instability scale $\mu_I$ towards the electroweak one, through the RG impact of their Yukawa coupling, provided their mass is low enough and their Yukawa couplings large enough. The question of whether the Majoron model can encompass these requirements is thus pertinent. The relevant fermionic Lagrangian in the Majoron model reads
\begin{align}
\mathcal{L}_{\nu_{R}} = \sum_{i} \overline{\nu_{R_i}} \slashed{\partial} \nu_{R_i}-\sum_{a, i} \boldsymbol{Y}_\nu^{a i} \bar{\ell}_L^a \tilde{H} \nu_{R_i} -\frac{1}{2} \sum_{i, j} \boldsymbol{Y}_R^{i j}\,S\,\overline{\nu_{R_i}^c} \nu_{R_j}+\text{h.c.}\,,
\label{eq:majoron:yukawa-couplings}
\end{align}
where $\boldsymbol{Y}_\nu$ and $\boldsymbol{Y}_R$ are Yukawa matrices and Latin indices run over the number of lepton doublets and right-handed neutrino species. Upon spontaneous symmetry breaking of $L$, see Eq.~(\ref{vevs}), a Majorana mass matrix for the $\nu_{R_i}$ fields is induced,
\begin{align}
    \boldsymbol{M_N} \equiv \frac{\boldsymbol{Y_R}}{2}\frac{w}{\sqrt{2}}\,,
      \label{eq:majoron:sterile-mass}
\end{align}
while the light neutrino mass matrix would take the form 
\begin{align}
    \boldsymbol{M_\nu} \equiv \frac{v^2}{\sqrt{2}}\, \boldsymbol{Y}_\nu \frac{1}{\boldsymbol{M_N}}\boldsymbol{Y}_\nu^T\,.
    \label{eq:light-neutrino-mass}
\end{align}
Given the tiny mass values of the observed neutrinos, this last equation could seem {\it a priori} incompatible with low BSM scales and sizeable neutrino Yukawa couplings (as welcome for a stringent bound on $m_h$~\cite{Khoury:2021zao}). A straightforward remedy is to consider low-scale Majoron models based on an approximate $U(1)_L$ symmetry~\cite{Branco:1988ex,Kersten:2007vk,Abada:2007ux,Moffat:2017feq}, as described in the Introduction. Popular realizations include the ``linear seesaw'' mechanism~\cite{Mohapatra:1986aw,Mohapatra:1986bd}, the ``inverse seesaw'' scenario~\cite{Akhmedov:1995ip,Malinsky:2005bi}, or the ``symmetry protected seesaw scenario'' (SPSS) in Ref.~\cite{Shaposhnikov:2006nn}. In the latter, the approximate symmetry enforces a vanishingly small value for the  combination of matrices  in Eq.~(\ref{eq:light-neutrino-mass}), while the individual entries in the $\boldsymbol{Y}_\nu$ matrix can be $\mathcal{O}(1)$. The main results of our paper will hold irrespective of the specific choice of low-scale Majoron model. In practice, for the numerical analyses we will use a version of the SPSS where the approximate $U(1)_L$ symmetry is displayed by the choice: 
\begin{align}
    \bld{Y}_R &= \left(
    \begin{array}{ccc}
    0 & 0   & 0   \\
    0 & 0   & Y_R \\
    0 & Y_R & 0
    \end{array}
    \right)\,, &
    \bld{Y}_\nu& =\left(\begin{array}{lll}
    0 & Y_{12} & 0 \\
    0 & Y_{22} & 0 \\
    0 & Y_{32} & 0
    \end{array}\right)\,,
\label{eq:majoron:SPSS-textures}
\end{align}
in which only $\nu_{R_{2,3}}$ couple to the SM and receive heavy degenerate masses, 
\begin{align}
M_N \equiv \frac{Y_R\,w}{2\sqrt{2}}\,,
\end{align}
while $\nu_{R_1}$ remains secluded as well as massless. This choice leads to very simple analytical expressions, because the RG analysis is only sensitive to the combinations~\cite{Chauhan:2023pur}
\begin{align}
\op{Tr}(\bld{Y}_{\nu}^\dagger\,\bld{Y}_{\nu}\,\bld{Y}_{\nu}^\dagger\,\bld{Y}_{\nu}) = \op{Tr}(\bld{Y}_{\nu}^\dagger\,\bld{Y}_{\nu})^2 = |Y_\nu|^4\,,
\label{eq:majoron:yukawa-combination}
\end{align}
where 
\begin{align}
|Y_\nu|^2 \equiv |Y_{12}|^2+|Y_{22}|^2+|Y_{32}|^2\,.
\label{eq:majoron:yukawa-trace-def}
\end{align}
In turn, the $\nu_{R_i}-\nu_{L_i}$'s mixing is then characterized by the following mixing angles:
\begin{align}
\Theta^a_\nu &=\frac{\bld{Y}_{\nu}^{* a 2}}{\sqrt{2}} \frac{v}{M_N}\,\quad \quad \text{and}   \quad\quad|\Theta_\nu|^2\equiv \sum_a\left|\Theta^a\right|^2\,,
\label{eq:majoron:mixing-angle}
\end{align}
where the index $a$ runs over the three lepton-doublet species.

\vspace{0.3cm}
\subsection{Input parameters: the Majoron scheme}
We will work in this article in the so-called ``Z-scheme'' for the SM electroweak bosonic sector. In this scheme, the four Lagrangian parameters $g$, $g'$, $\mu_H$ and $\lambda_H$ are traded for four measurable input parameters: $\alpha_\text{em}$ denoting the fine-structure constant as measured from Thompson scattering, the Fermi constant $G_F$ as extracted from muon decay, and $m_Z$ and $m_h$ denoting the pole values of the $Z$ and Higgs mass, respectively.

The Majoron model discussed above adds five relevant free parameters: $\mu_S$, ${Y}_R$, $\lambda_S$, $\kappa$ and  ${Y}_\nu$, see Eqs.~(\ref{eq:majoron:potential}) and (\ref{eq:majoron:sterile-mass}). When considering measurable quantities further below we will trade the first two for $M_N$ and $M_s$. In summary, we will work in a scheme in which the physical independent quantities will be expressed in terms of the set 
\begin{equation}
\{ \underbrace{\alpha_\text{em}, G_F, m_Z, m_h}_{\text{Z-scheme}}, \underbrace{M_N, M_s, \kappa, Y_\nu, \lambda_S}_{\text{Majoron model}}\}\,.
\label{eq:Majoron-scheme}
\end{equation}
We will refer to this ensemble as {\it Majoron scheme}. Nevertheless, in some illustrations we will replace $Y_\nu$ by the light-heavy neutrino mixing $\Theta_\nu$, see Eq.~(\ref{eq:majoron:mixing-angle}), as this parameter is often used to explore experimental constraints. Precise relations between the input parameters above and the Lagrangian parameters can be found in \cref{sec:appendix:input-scheme}.

\subsection{Effective field theory}
\label{sec:EFT}
For energies $M_N<E<M_s$, the effective Lagrangian appropriate to confront collider data reads
\begin{align}
\mathcal{L}_{\nu\text{SMEFT}} = \mathcal{L}_{SM}+
\mathcal{L}_{\text{seesaw}} 
+ \mathcal{L}_{\text{eff}}
\,,
\label{Lag}
\end{align}
where $\mathcal{L}_{\text{seesaw}}$ describes the type I seesaw Lagrangian for heavy neutrinos which results after integrating out the radial scalar field $S$ in \cref{eq:majoron:yukawa-couplings}\,, 
\begin{align}
\mathcal{L}_{\text{seesaw}} = \sum_{i} \overline{\nu_{R_i}} \slashed{\partial} \nu_{R_i}-\sum_{a, i} \boldsymbol{Y}_\nu^{a i} \bar{\ell}_L^a \tilde{H} \nu_{R_i} - \sum_{i, j} \boldsymbol{M_N}^{i,j}\,\overline{\nu_{R_i}^c} \nu_{R_j}+\text{h.c.}\,.
\label{eq:seesaw}
\end{align}
The effective Lagrangian $\mathcal{L}_{\text{eff}}$ contains  mass dimension $d=5$ and $d=6$ couplings. Its $d=5$ operators violate lepton number, while the $d=6$ set includes SMEFT operators as well as others containing the $N$ fields,
\begin{align}
    \mathcal{L}_{\text{eff}} =  \frac{1}{\Lambda}\sum_{i} C_k\,\mathcal{O}^{d = 5}_{k}+\frac{1}{\Lambda^2}\sum_{i} C_k\,\mathcal{O}^{d = 6}_{k}\,,
    \label{Leff}
\end{align}
where $\Lambda$ is the high-energy physics scale, here identified with the heavy scalar mass $\Lambda=M_s$, and $C_i$ denote arbitrary Wilson coefficients. A convenient basis for the SMEFT couplings is the {\it Warsaw basis}~\cite{Grzadkowski:2010es}, in which three of its $d=6$ operators contribute to the Higgs potential, 
\begin{align}
\mathcal{L}_{\text{eff}}\supset \frac{C_H}{\Lambda^2}\,\OO_{H} + \frac{C_{H\square}}{\Lambda^2}\,\OO_{H\square} + \frac{C_{H D}}{\Lambda^2} \OO_{H D}\,,
\label{eq:operators-definition}
\end{align}
where
\begin{align*}
&\OO_H \equiv (H^\dagger H)^3\,,
&&\OO_{H \square} \equiv (H^\dagger H)\square (H^\dagger H)
&&\text{and}
&&\OO_{H\,D} \equiv (H^\dagger D^\mu H)^*(H^\dagger D_\mu H)\,.
\label{eq:SMEFT-operators}
\end{align*}
\begin{table}[h!]
\begin{align*}
\begin{array}{ll}
\hline\hline
\mathcal{O}_H=(H^{\dagger} H)^3 & \mathcal{O}_{H W}=(H^{\dagger} H) W_{\mu \nu}^a W^{a \mu \nu} \\
\mathcal{O}_{H \square}=(H^{\dagger} H) \square(H^{\dagger} H) & \mathcal{O}_{H B}=(H^{\dagger} H) B_{\mu \nu} B^{\mu \nu} \\
\mathcal{O}_{H D}=(H^{\dagger} D^\mu H)^*(H^{\dagger} D_\mu H) & \mathcal{O}_{H W B}=(H^{\dagger} \sigma^a H) W_{\mu \nu}^a B^{\mu \nu} \\
\mathcal{O}_{u H}=(H^{\dagger} H)(\bar{q} u \widetilde{H}) & \mathcal{O}_{H e}=(H^{\dagger} i \stackrel{\leftrightarrow}{D} \mu H)(\bar{e} \gamma^\mu e) \\
\mathcal{O}_{d H}=(H^{\dagger} H)(\bar{q} d H) & \mathcal{O}_{H q}^{(1)}=(H^{\dagger} i \stackrel{\leftrightarrow}{D_\mu} H)(\bar{q} \gamma^\mu q) \\
\mathcal{O}_{e H}=(H^{\dagger} H)(\bar{\ell} e H) & \mathcal{O}_{H q}^{(3)}=(H^{\dagger} i \stackrel{\leftrightarrow}{D_\mu^a} H)(\bar{q} \gamma^\mu \sigma^a q) \\
\mathcal{O}_{H u}=(H^{\dagger} i \stackrel{\leftrightarrow}{D_\mu} H)(\bar{u} \gamma^\mu u) & \mathcal{O}_{H \ell}^{(1)}=(H^{\dagger} i \stackrel{\leftrightarrow}{D_\mu} H)(\bar{\ell} \gamma^\mu \ell) \\
\mathcal{O}_{H d}=(H^{\dagger} i \stackrel{\leftrightarrow}{D} \mu H)(\bar{d} \gamma^\mu d) & \mathcal{O}_{H \ell}^{(3)}=(H^{\dagger} i \stackrel{\leftrightarrow}{D_\mu^a} H)(\bar{\ell} \gamma^\mu \sigma^a \ell)\\
\hline\hline
\end{array}
\end{align*}
\caption{SMEFT operators of $d=6$ in the Warsaw basis generated by integrating out the heavy singlet scalar at one loop. The $\sigma^a$ are the Pauli matrices, and $\stackrel{\leftrightarrow}{D}{}_\mu^a \equiv \stackrel{\leftrightarrow}{D}_\mu \sigma^a$.}
\label{tab:Warsaw-operators-SMEFT}
\end{table}

\noindent The relevance of  $\OO_H $ for the Higgs potential is straightforward once $H$ develops a VEV. The other two operators include derivatives, but nevertheless they modify the scalar potential upon canonical normalization of the Higgs field, whose kinetic energy they modify -- see~\cref{sec:appendix:input-scheme}. 
For the Majoron model and at tree-level,  $\OO_{H \square}$ is generated from \cref{eq:majoron:potential} after integrating out $S$, 
\begin{align}
    C_H^{(0)} &= 0\,, &  C_{H\square}^{(0)} &= -\frac{\kappa^2}{4\,\lambda_S}\,,&C_{H\,D}^{(0)}= 0\,,
\end{align}
where the superscript $(0)$ signals  tree-level quantities. In addition, the term quadratic term |H| and the Higgs self-coupling in Eq.~\eqref{eq:majoron:potential}
also receive tree-level corrections:
\begin{align}
 \mu_H^2 \to  \mu_H^2- \frac{\kappa}{2\lambda_S}\,, \qquad \qquad \lambda_H \to\lambda\equiv \lambda_H -\frac{\kappa^2}{4\lambda_S}\,.
 \label{matching-tree-level}
\end{align}
After canonical normalization of the Higgs kinetic energy to include the impact of $C_{H\square}^{(0)} $, the tree-level Higgs potential reads
\begin{align}
V(h) 
&= h^2 \,{\lambda} \,v^2 \left(1 + 2\,\frac{v^2}{M_s^2}\,C_{H\square}^{(0)}\right)
+ h^3\, {\lambda}\,v \left(1 + 5\,\frac{v^2}{M_s^2}\,C_{H\square}^{(0)}\right)+\nn\\&\hspace{7.5cm} + h^4 \,\frac{{\lambda}}{4} \left( 1 + \frac{56}{3}\,\frac{v^2}{M_s^2}\, C_{H\square}^{(0)}\right)\,,
\label{Vhtree}
\end{align}
where $v$ is as defined in Eq.~(\ref{vevs}), with either $v^2/2=  (2\lambda_S \mu_H^2-\kappa\mu_S^2)/(4 \lambda_H \lambda_S-\kappa^2)$ for a non-trivial Higgs minimum, or $v=0$ as an extremum.  In the Z-scheme -- \cref{eq:Majoron-scheme} -- this tree-level potential takes the form:
\begin{align}
V(h) &= 
  \frac{m_h^2}{2}\,h^2
+ \frac{m_h^2G_F^{1/2}}{2^{3/4}}\,g_{hhh}\,h^3
+ \frac{m_h^2\,G_F}{4\sqrt{2}}\,g_{hhhh}\,h^4\,,
\end{align}
where 
\begin{equation}
m_h^2= 2 \,{\lambda} \,v^2 \left(1 + 2\,\frac{v^2}{M_s^2}\,C_{H\square}^{(0)}\right)\,, \qquad \qquad G_F= \frac{1}{\sqrt{2}\, v^2},
\label{mhtree}
\end{equation}
and 
\begin{align}
g_{hhh} = 1  + \frac{3}{\sqrt{2} }\,\frac{1}{G_F\,M_s^2}\, C_{H\square}^{(0)}
\,\,,\qquad\qquad\qquad
g_{hhhh} = 1 +
\frac{50}{3\sqrt{2}}\,\frac{1}{G_F\,M_s^2}\, C_{H\square}^{(0)}
\,\,.
\end{align}
These $C_{H\square}$ corrections will dominate the impact on the cubic and quartic Higgs self-coupling to be explored in (present and future) accelerator data, as discussed in the next subsection. Note that the dominance of $\OO_{H \square}$ differs from previous analyses of Higgs near-criticality using pure EFT, which focused instead on $\OO_H $.
 
The one-loop impact of the Majoron model on the SMEFT coefficients has been computed in the literature~\cite{Jiang:2018pbd}: then, the operators $\OO_{H}$ and $\OO_{HD}$ start  to play a role in the scalar potential and contributions to the coefficients of other $d=6$ operators involving the Higgs field and SM field strengths or fermions become relevant as well. The SMEFT operators involved are displayed in \cref{tab:Warsaw-operators-SMEFT} while their one-loop expressions for the Wilson coefficients  can be found in \cref{sec:appendix:SMEFT-scalar}, for the particular case of the Majoron model. The future sensitivity of the HL-LHC and the FCC-ee~\cite{Celada:2024mcf} project to those SMEFT coefficients has been recently assessed, enabling bounds on most coefficients of the operators in \cref{tab:Warsaw-operators-SMEFT}, and in particular on $C_{DH}$, although not on $C_{H}$. Sensitivity to the latter will be better achieved at the FCC-hh~\cite{Mangano:2020sao} through the Higgs trilinear coupling $g_{hhh}$, which may be constrained at the percent level. The quantitative analysis will be developed in \cref{sec:experimental-limits}.   

Besides the SMEFT operators discussed, the effective Lagrangian in Eq.~(\ref{Leff}) contains couplings involving one or more neutrino fields $N_i$~\footnote{We are indebted to P. P. Giardino for pointing this out.}: the ensemble  is often denominated $\nu$SMEFT~\cite{Graesser:2007yj,delAguila:2008ir,Aparici:2009fh}. In the model under discussion and at tree level,  a $d=5$  coupling is generated: 
\begin{align}
\bld{\OO}_{N H}  =\bar{N}_i N_j^c H^{\dagger} H +h . c.\,, \qquad \text{with} \qquad \frac{\bld{C}^{(0)}_{NH}}{\Lambda} = -2\sqrt{2}\,\kappa \, \frac{\bld{M}_N}{M_s^2}\,,
\end{align}
as well as a $d=6$ operator:  
\begin{align}
\bld{\OO}_{NN} = (\bar{N} N^c)(\bar{N} N^c)\,, \qquad \text{with} \qquad \frac{\bld{C}^{(0)}_{NN}}{\Lambda^2} = 
24 \,\lambda_S\, \frac{\bld{M}_N^2}{M_s^4}\,, 
\end{align}
where four flavour indices in the last equation have been left implicit, and the bold typography indicates that these are matrices in flavour space. The coefficients of these two operators are strongly suppressed by powers of $M_N/M_s$, but this could be {\it a priori} comparable to one-loop radiative corrections. Nevertheless, these operators have been analyzed in the literature~\cite{Caputo:2017pit,Fernandez-Martinez:2023phj} and turn out to have subleading effects in our case. For instance  Ref.~\cite{Fernandez-Martinez:2023phj} showed that $\OO_{N H}$ will impose significant limits on the parameter space for heavy neutrinos only in the mass range $M_N<M_Z$, and in consequence   not  for our region of interest which spans  $M_N\ge 1$ TeV. The impact of $\OO_{NN}$ is even further beyond any foreseable collider prospect.

\subsection{Experimental limits and prospects at future colliders for the Majoron model
}
\label{sec:experimental-limits}
The metastability bound on $m_h$ is of about the same order of magnitude as  the instability scale $\mu_I$, see Eq.~(\ref{eq:metastability-bound}), and in consequence the requirement of a tight bound on $m_h$ brings down all scales of the problem towards the  $\#$TeV range, see Eq.~(\ref{scales}).  This implies that present and future electroweak and collider data become relevant to test that metastability paradigm.  In particular, the FCC energy range may allow to directly sweep over a good fraction of the parameter space of the  low-scale Majoron model. This compounds the imperious need to analyse the metastability predictions for the Higgs mass  in the light of present and future collider data.
 
The RGE of the Higgs self-coupling plays a fundamental role in the present study. In the Standard Model of Particle Physics (SM),  $\lambda$ corresponds to the third derivative of the Higgs potential, which at tree-level is given by $\lambda_{SM}= m_h^2/(2v^2)$, where $v$ is independently determined from electroweak measurements. This results in a SM tree-level prediction $\lambda_{SM}\simeq 0.129$~\cite{ATLAS:2023oaq}, which is crucial to test. HL-LHC will tackle it with about $~\mathcal{O}(50\%)$ accuracy at $68\%$ C. L., while an $~\mathcal{O}(20\%)$ accuracy will be needed for a $5\sigma$ discovery of $\lambda$ (assuming a Gaussian likelihood function). A future accelerator such as the FCC-ee could improve the accuracy to $~\mathcal{O}(24\%)$ (with 4 interaction points~\cite{deBlas:2019rxi}), while the FCC-hh could reach $~\mathcal{O}(5\%)$, entering the quantum regime. Furthermore, the FCC-ee will already provide percent-level accuracy for other interactions, e.g. for many SMEFT effective couplings. Overall, to achieve percent-level accuracy -- as the FCC project could enable -- it is tantamount to enter the quantum regime, essential to understand the nature of the Higgs potential and in particular naturalness issues. For the case under consideration in this work, this is precisely the appropriate level of accuracy to confront the loop-level impact of the Majoron model in the low-mass range relevant for a stringent metastability bound on the Higgs mass. 

Given the energy scales at play for the low-scale Majoron model, the heavy Majorana neutrinos will be dynamical down to the FCC-ee energies, while the exotic scalar sector will be outside direct FCC reach. The exotic scalar sector can thus be integrated out for the purpose of analyzing collider data, and EFT methods will be appropriate to explore its impact on the FCC-hh data, see Eqs.~(\ref{Lag})-(\ref{Leff}) and the detailed analysis in \cref{sec:EFT}. 

  \subsubsection*{Neutrino sector}

\begin{figure}[h!]
    \centering
    \includegraphics[width = 0.49\textwidth]{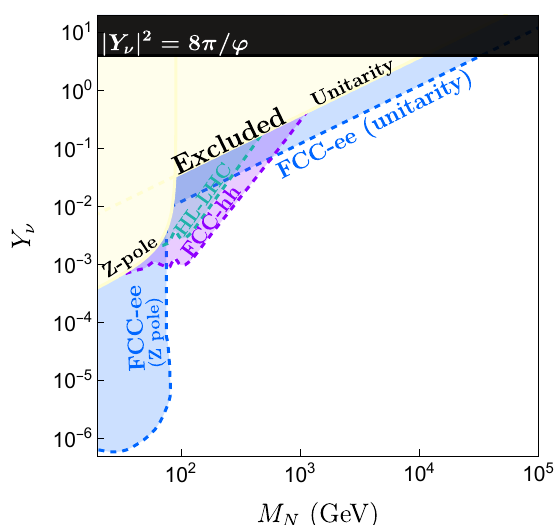}
    \includegraphics[width = 0.49\textwidth]{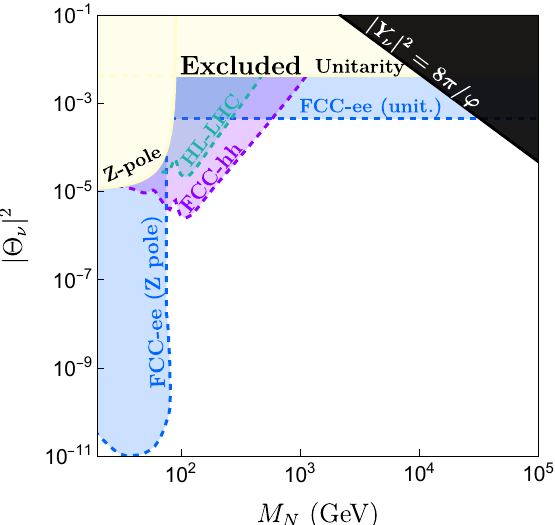}
    \caption{
    Current and prospected collider limits on the HNL parameter space in terms of the mass of the neutrinos $M_N$  vs. the Yukawa coupling $Y_\nu$ (left) and the mixing angle $|\Theta|$ (right). 
    }
    \label{fig:neutrino-limits-prospects}
\end{figure}

Experimental limits and prospects on the parameter space of the heavy neutrino sector are displayed in \cref{fig:neutrino-limits-prospects} in terms of the mass of the heavy neutrino $M_N$ vs. the total mixing angle $|\Theta|^2$ (left) and the Yukawa coupling $Y_\nu$ (right). The latter two parameters are related by \cref{eq:majoron:mixing-angle}. Excluded regions are displayed in yellow. For $M_N < M_Z$, direct searches at collider experiments~\cite{DELPHI:1996qcc,ATLAS:2015gtp,CMS:2018iaf,CMS:2018jxx,ATLAS:2019kpx,LHCb:2020wxx,CMS:2022fut,ATLAS:2022atq,ATLAS:2023tkz,CMS:2024ake} provide the strongest constraints; we display some of the most stringent results from the Z-pole search by DELPHI at Lep-I~\cite{DELPHI:1996qcc}. Direct searches are not effective for larger HNL masses. However, their presence will produce deviations from unitarity of the PMNS leptonic mixing matrix. The searches of non-unitary mixing are dominated by the non-detection of charged lepton flavor violating (cLFV) processes such as $\mu\rightarrow e\gamma$, $\tau\rightarrow \ell\,\gamma$ or $\mu \rightarrow eee$, but include as well limits on electroweak precision observables (EWPO) and universality ratios. Plenty of analyses of these deviations exist in the literature -- e.g. Refs.~\cite{Antusch:2014woa,Antusch:2015mia,Blennow:2023mqx} and references therein. In our figure, we display the up-to-date results from the global analysis in Ref.~\cite{Blennow:2023mqx} and label that region as ``Excluded (Unitarity)''.

Theoretical limits on the heavy neutrino parameter space also exist~\cite{Urquia-Calderon:2024rzc}, stemming from the preservation of perturbative unitarity, depicted in black in \cref{fig:neutrino-limits-prospects}. The latter become relevant for larger heavy neutrino masses, the strongest being established by $J = 1$ amplitudes, setting $|Y_\nu|^2 < 8\,\pi/\varphi$, where $\varphi = (1+\sqrt{5})/2 \simeq 1.618\dots$ is the golden ratio.

The predictions at future colliders (in green, blue and mauve) for a type-I seesaw compatible with the SPSS low-scale Majoron are taken from  Ref.~\cite{Antusch:2016ejd}. Other useful sources including similar prospects can be found in Ref.~\cite{Beacham:2019nyx}. Some of the limits reported in those references are flavor-sensitive -- they constrain the partial mixing angles $|\Theta_a|$ -- while we are interested in the flavor-insensitive ``total'' mixing angle $|\Theta|$. In those cases we consider the weakest limit on the $|\Theta_a|$, which is typically $|\Theta_\tau|$. This approach is optimistic in that it maximizes the available parameter space, but conservative for the detection potential at future colliders, since it underestimates their detection capability for heavy neutrinos with larger $|\Theta_e|$ and $|\Theta_\mu|$ mixings.

\subsubsection*{Scalar sector}
\begin{figure}[h!]
    \centering
    \includegraphics[width = 0.49\textwidth]{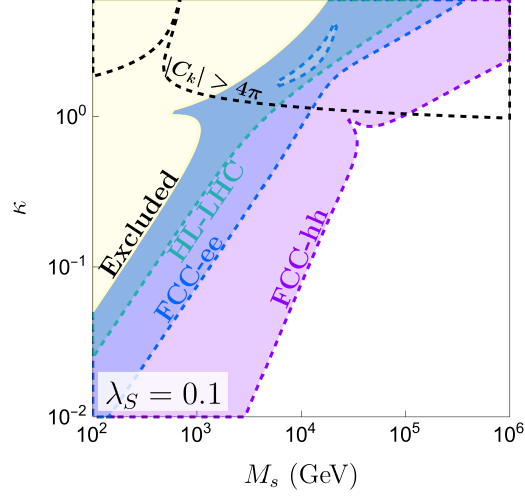}
    \includegraphics[width = 0.49\textwidth]{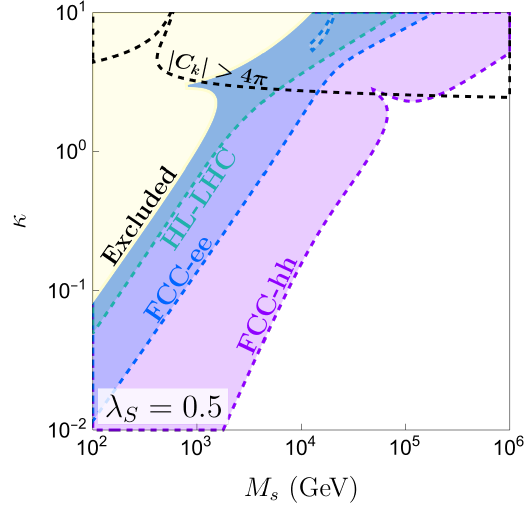}
    \caption{
    Current and prospected collider limits on the exotic heavy scalar sector of the Majoron model for $\lambda_S = 0.1$ (left) and $\lambda_S = 0.5$ (right). Above the black dashed line, one or more of the effective couplings in the SMEFT may become non-perturbative. 
    }
    \label{fig:scalar-limits-prospects}
\end{figure}

Given the hierarchy of scales in Eq.~(\ref{scales}) with $M_N < M_s$, and that the neutrino sector must dominate the RG evolution long enough to sizably lower the instability scale $\mu_I$, the heavy scalar field $S$ will not appear as an active degree of freedom in any of the colliders considered in this work. This permits  an EFT approach to the scalar sector impact on collider data. For LHC, HL-LHC and FCC-ee, we combine the SMEFT and $\nu$SMEFT analysis in \cref{sec:EFT} with the global fit recently presented in~Ref.\,\cite{Celada:2024mcf}. While we  consider the whole set of SMEFT operators generated by integrating out $S$ at the tree and one-loop levels -- given in \cref{tab:Warsaw-operators-SMEFT} -- the limits are dominated by far by the contribution of the $\OO_{H\square}$ operator, which is the only one arising at  tree-level. For the FCC-hh prospects, the results in Ref.\,\cite{Mangano:2020sao} are used instead, indicating a prospective measurement of $\kappa_\lambda \equiv \lambda/\lambda_{SM}$ with up to $2.8\%$ accuracy. The resulting limits and prospects are displayed in \cref{fig:scalar-limits-prospects}. The regions above the ``$|C_k| > 4\pi$'' lines are to be considered excluded as some of the parameters lie outside the perturbative regime. 

\section{Metastability bound on the Higgs mass}
\label{sec:2-field-potential-near-criticality}

The metastability bound on the Higgs mass~\eqref{eq:metastability-bound} was first found in Ref.~\cite{Buttazzo:2013uya}, and completed in Refs.~\cite{Khoury:2021zao,Benevedes:2024tdq}.  
The general conditions for the existence of such a  bound are that:
\begin{enumerate}[label=\textbf{\arabic*}.]
\item The EW vacuum does not correspond to a stable minimum but to a metastable one.
\item The EW vacuum allows for the EWSB pattern of the SM.\footnote{A 
 metastability bound has recently been found in Ref.~\cite{Benevedes:2024tdq} for the case of a positive Higgs mass parameter in the low-energy theory (although EWSB does not take place in this case). This bound can be even more stringent than the one for a negative mass parameter. This implies that a negative mass term in the Higgs potential is not essential for the existence of a metastability bound.
 } 
\item The vacuum into which the Higgs field tunnels with the highest probability requires the Higgs self-coupling $\lambda$ to become negative due to its RG running.
\end{enumerate} 
The ability of the metastability bound to explain the smallness of the observed Higgs mass hinges on a lowering of the instability scale, which a series of previous works~\cite{Giudice:2021viw,Khoury:2021grg,Benevedes:2024tdq} 
suggested to achieve through the addition of BSM fermions.\footnote{For an alternative approach using axion-like particles (ALPs) instead, see Ref.~\cite{Detering:2024vxs}.} Doing so would destabilize the vacuum, though, to the point where its expected lifetime would be smaller than its current age. This led to consider in addition the stabilizing effects stemming from BSM scalar physics. The latter have been included in those previous works in a rather model-independent way through the addition of a dimension-six scalar SMEFT  effective operator (i.e., built out of the Higgs field). While this picture is indeed sufficient to explore many important aspects of the metastability bounds, it is linked to three conceptual problems: i) the bound requires a very specific hierarchy of scales to be assumed, $m_{\rm fermion} \lesssim \mu_I \lesssim {\Lambda}$, where $m_{\rm fermion}$ denotes the mass of the BSM fermions and $\Lambda$ is the EFT generic scale of new physics; ii) in the regimes in which the bound is the most successful, i.e., most stringent and close to the observed value for $m_h$, a reliable calculation of the lifetime would require to include higher-dimensional EFT operators;  iii)  the condition {\bf 3} above had to be explicitly imposed.

\subsection{Metastability bound and quantum phases: the Majoron model}\label{sec:MSBoundPhases}
To address those three  open questions at once, we consider the realization of the Higgs mass metastability bound in a concrete model -- the low-scale Majoron model reviewed in \cref{sec:majoron-completion}. 
 
The new physics scale can then be identified with the heavy singlet scalar  mass $\Lambda=M_s$, and  condition {\bf 2} above implies that {\it in the false vacuum} both scalar fields develop a non-vanishing VEV due to a negative Higgs mass term, $|\langle H \rangle|=v/\sqrt{2}$ and $\langle S \rangle =w/\sqrt{2}$, with $v\ll w$. In turn, condition {\bf 1} singles out as viable scenarios from the landscape in \cref{fig:potential:configurations} the configurations in which our universe is associated -- at tree level -- to the $\ovbld{P}_2$ saddle point, and a separate minimum is also present. That is, phases \tbf{II}, \tbf{II'} and \tbf{IV}. Furthermore, condition {\bf 3}  is an automatic consequence of this Majoron model setup.

The effective theory of the Higgs in the vicinity of the saddle point (where the Higgs VEV is negligible compared to that for $S$) can be linked to the full theory through the standard matching procedure, see the tree-level matching condition in Eq.~(\ref{matching-tree-level}). This equation shows that the requirements of a small Higgs mass and a negative quartic coupling in the intermediate energy region imply
\begin{gather}\label{eq:matchingparameters}
     \mu_H^2 \gtrsim \frac{\kappa}{2 \lambda_S} \mu_S^2 \ \ \quad \text{and} \ \quad\kappa^2 > 4 \lambda_H \lambda_S .
\end{gather}
In terms of the parameters $\Omega_H$ and $\Omega_S$, this corresponds to
\begin{align}
    \Omega_H \lesssim 4  \frac{\lambda_S \lambda_H}{\kappa^2}<1\,, \quad  \ \  \ \Omega_S \gtrsim 1\, \quad \ \ \ \ \text{and} \quad \kappa>0\,,
    \label{metaconditions}
\end{align}
see Eqs.~(\ref{eq:potential:definition-Omega_H-Omega_S})-(\ref{eq:potential:global-stability-conditions-negative}). The conditions in Eq.~(\ref{metaconditions}) eliminate two of the three vacuum configurations identified as {\it a priori} suitable in Sec.~\ref{sec:majoron-completion}: the first and third condition are not simultaneously satisfied by phase \tbf{II}, while the last one is not satisfied by phase \tbf{II}', see Fig.~\ref{fig:potential:configurations}. This leaves \tbf{IV} as the optimal tree-level configuration. 

The picture changes  if RG effects on the potential are taken into account. We will show in the next subsection that, for a small enough mass parameter, radiative corrections can lead to the formation of an additional -- metastable -- vacuum in the vicinity of a saddle point of the tree-level potential. Conversely, demanding the formation of such a vacuum generally requires a small Higgs mass. 

\subsection{Radiatively generated vacua and metastability bound}
\label{sec:RGV}

\subsubsection*{The metastability bound in the Majoron model}
\begin{figure}[h!]
    \centering
    \begin{minipage}{0.49\textwidth}
\includegraphics[width=\textwidth]{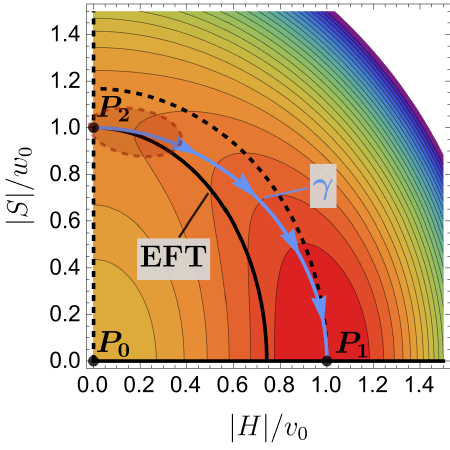}
    \end{minipage}
    \begin{minipage}{0.465\textwidth}
    \vspace*{-0.3cm}
    \includegraphics[width=\textwidth]{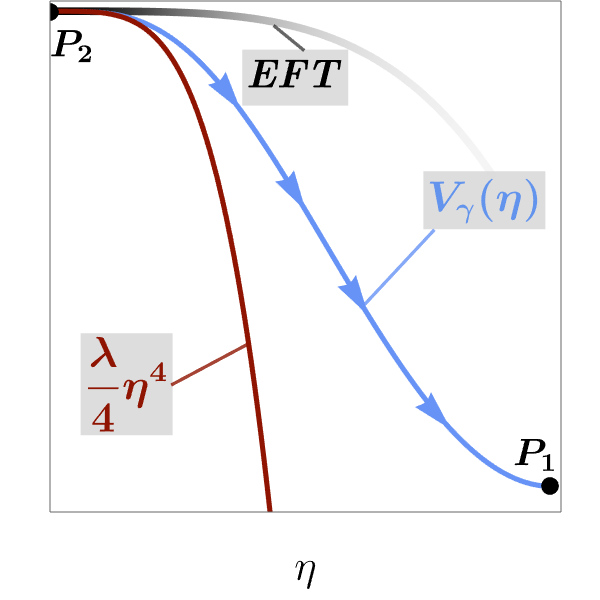}
    \end{minipage}
    \caption{\tbf{Left:} example of configuration in Phase \tbf{IV}. The steepest-descent contour $\gamma(\eta)$ -- along which $V_\gamma(\eta)$ is defined -- is shown in blue.  Meanwhile, the  
    black solid line represents the EFT described in
    \cref{sec:majoron-completion}, obtained by substituting $|S|$ by its $H$-dependent VEV as defined in Eq.~(\ref{Sminima}). In the neighborhood of $\bld{P_2}$, highlighted as a reddish oval area, both theories coincide. \textbf{Right:} tree-level profile of $V_\gamma(\eta)$, depicting only its quartic term. The quartic term in $V_\gamma(\eta)$, with $\lambda < 0$, is stabilized by the extra terms in $F_{2n}(\eta/M_s)$ as defined in \cref{eq:V-gamma}.
   }    \label{fig:tunneling:potential+gradient-descent}
\end{figure}

Having identified the part of parameter space where the generation of an additional vacuum is in principle feasible, we can now proceed to analyze this mechanism in full detail. 
In order to do so, let us consider the vicinity of one of the saddle points in the tree-level potential. Without the potential barrier, the field would simply roll down the potential along the steepest descent contour 
\begin{align}
\gamma (\eta)\equiv(|H|_\gamma (\eta), |S|_\gamma (\eta))
\end{align}
connecting the saddle point to the adjacent lower-lying vacuum, where $\gamma(\eta)$ describes the trajectory in the two-dimensional field space $\{|H|, |S|\}$ as a function of the location parameter $\eta$.  This contour only agrees with the usual effective theory corresponding to simply integrating out the heavy field $S$ for very small values of $|H|$. An example is depicted in blue in \cref{fig:tunneling:potential+gradient-descent}, which illustrates a transition from the saddle point $P_2$ to the minimum $P_1$, while the naive EFT line corresponding to an expansion around $P_2$ would have resulted in the solid black trajectory (an analogous argument applies to the dotted black line stemming from $P_1$ and corresponding to negligible $S$ VEVs - $w$ - and large $H$ VEVs - $v$). 

It is possible to parameterize $\gamma$ in such a way that $\eta$ can be interpreted as a canonically normalized field itself, with Lagrangian 
\begin{align}
    \mathcal{L}_\gamma= \frac{1}{2}\partial_\mu \eta \partial^\mu \eta - V_{\gamma} (\eta), \ \ \text{where}\ V_\gamma (\eta) = V(H_\gamma (\eta),S_\gamma (\eta))\,,
\end{align}
where in this equation and what follows we simplify the notation identifying 
$H_\gamma(\eta)\equiv |H|_\gamma(\eta)\,, \,\,S_\gamma(\eta)\equiv |S|_\gamma(\eta)$. This parametrization can be implemented through the field redefinition
\begin{align}
    \frac{{\rm d}}{{\rm d} \eta} 
    \begin{pmatrix}
    H_\gamma(\eta) \\
    S_\gamma(\eta)
\end{pmatrix}
     = \frac{- \nabla_{\{H, S\}} V(H_\gamma (\eta), S_\gamma (\eta))}{\sqrt{2}|\nabla_{\{H, S\}}  V(H_\gamma (\eta), S_\gamma (\eta))|}\,.
     \label{eq:tunneling:gradient-descent}
\end{align}
The formation of the {\it radiatively generated vacuum} thus requires the formation of a barrier in the effective potential obtained by evaluating the full potential~\cref{eq:majoron:potential} along the contour $\eta$. It is straightforward to solve Eq.~\eqref{eq:tunneling:gradient-descent} numerically.

Nevertheless, in order to gain physical insights and to relate the parameters in $V_\gamma$ to those in the SM potential we can analyze the contour $\gamma$ for small values of $\eta$, i.e., near the saddle point $\bld{P_2}$ in \cref{eq:V-gamma}, which later on will correspond to the EW vacuum. For small $\eta$, this contour is of the general form 
\begin{align}\label{eq:gammapprox}
    \begin{pmatrix}
    H_\gamma (\eta) \\
    S_\gamma (\eta)
\end{pmatrix}
     = \frac{1}{\sqrt{2}} \begin{pmatrix}
     \eta \left[ 1 - H_2 \frac{\eta^2}{M_s^2} + ... \right] \\
    w \left[1- S_2  \frac{\eta^2}{M_s^2}  ...\right]
\end{pmatrix}\,,
\end{align}
where $H_2, S_2,...$ are functions of the potential's parameters.  
Due to this scaling, we find that the potential in terms of $\eta$ takes the familiar form
\begin{gather}
    V_\gamma (\eta)= - \frac{\mu_\eta^2}{2}\eta^2 + \frac{\lambda_\eta}{4} \eta^4  + \eta^4 F \left( \frac{\eta}{M_s} \right), \ \  \label{eq:V-gamma} \\ 
    \text{with} \qquad  F \left( \frac{\eta}{M_s} \right) = \sum_{n=1}^{\infty}  F_{2n} \left( \frac{\eta}{M_s} \right)^{2n}\,.
    \label{eq:V-gamma-2}
\end{gather}
The parameter $\mu_\eta^2$ is related to the tree-level Higgs mass $m_h^2$ after spontaneous symmetry breaking through $m_h^2=2\mu_\eta^2$ at tree level, as in \cref{mhtree}. For simplicity, we will report our bound directly in terms of $m_h^2$. In the limit $m_h^2 \ll M_s^2$ the parameter $\lambda_\eta$ can be identified with the effective quartic coupling of the tree-level low-energy Higgs theory at the matching scale $\mu = M_s$.

We will move forward using consistently $\lambda$ instead of $\lambda_\eta$, as that strong hierarchy between the Higgs and S masses will ultimately be mandated by the the metastability bound. The function $F$ is an infinite polynomial series of even powers of $\eta$ reflecting the initial $Z_2$ symmetry, and $F_{2n}$ are polynomial coefficients. These have an explicit dependence on the parameters of the scalar sector in the tree-level potential, i.e., $\lambda_H, \lambda_S$ and $\kappa$. While in our Majoron scheme $\lambda_S$ and $\kappa$ are indeed independent parameters -- see Eq.(\ref{eq:Majoron-scheme}), we can treat $\lambda_H$ as set by Eq.~\eqref{matching-tree-level}. The value of $\lambda$ in this equation depends not only on $M_S$ and the SM input for the relevant couplings, but also on $M_N$ and $Y_\nu$ when one-loop effects are considered in addition, as they generically have a significant impact on the running of $\lambda$. This also establishes a correlation between the parameters in $F$ and the instability scale $\mu_I$, whose value is also set by $M_N$ and $Y_\nu$. 

In order to obtain Eqs.~\eqref{eq:gammapprox} and~\eqref{eq:V-gamma}, we have expanded $\gamma$ for small $\eta$, making these expressions only appropriate in the vicinity of the saddle point $\bld{P_2}$. Throughout our analysis, however, we use the full expression obtained by numerically integrating Eq.~\eqref{eq:tunneling:gradient-descent}. Notably, this equation treats $H$ and $S$ on an equal footing, in principle allowing for the construction of a similar effective description around $\bld{P_1}$. Denoting by $\eta_{1}$ the value of $\eta$ for which $\gamma (\eta_{1})=\bld{P_1}$, we find that in this regime $s\sim  \eta_1 - \eta$, as suggested by Fig.~\ref{fig:tunneling:potential+gradient-descent}. Moreover, as an additional cross-check of our steepest-descent code used to solve Eq.~\eqref{eq:tunneling:gradient-descent}, we explicitly check that $\gamma$ indeed reaches $\bld{P_1}$. 

Following Sec.~\ref{sec:MSBoundPhases}, the potential in Eq.~\eqref{eq:V-gamma} does not give rise to a metastable vacuum at tree level. To incorporate the quantum corrections necessary to allow for the formation of such a vacuum, we therefore include loop corrections to the coupling $\lambda$. For $\eta < M_S$, we therefore replace the tree-level coupling $\lambda$ in Eq.~\eqref{eq:V-gamma} by its full one-loop expression, 
\begin{align}
    \lambda \to \lambda(\mu, \eta) \equiv \lambda (\mu)+ \delta \lambda(\eta,\mu),
    \label{lambda_mu_eta}
\end{align}
where $\delta \lambda(\mu, \eta)$ is given in Refs.~\cite{Andreassen:2014gha,Chauhan:2023pur}, 
\begin{align}
    (4 \pi)^2\delta \lambda (\mu, \eta)=& -\frac{15 g^4}{32} -\frac{5 g^2 (g^{\prime})^2}{16}-\frac{5 (g^{\prime})^4}{32}+\frac{9 y_t^4}{2}+\frac{3}{8} g^4 \log \left(\frac{g^2}{4}\frac{\eta^2}{\mu ^2}\right) \nonumber \\    
    & +\frac{3}{16} \left(g^2 + (g^{\prime})^2 \right)^2 \log \left(\frac{g^2+(g^{\prime})^2}{4}\frac{\eta ^2}{\mu ^2}\right)-3 y_t^4 \log \left(\frac{y_t^2}{2}\frac{\eta^2}{\mu^2}\right) \nonumber \\ 
    &+\frac{3 |Y_\nu|^4}{2}-  |Y_\nu|^4 \log \left(\frac{|Y_\nu|^2}{2}\frac{\eta^2}{\mu^2}\right)\,.
    \label{eq:deltalambda}
\end{align}
The first two lines collect the contributions arising from the Higgs' couplings to the SM particle content, which we take into account for all scales $\mu < M_S$. The terms in the third line arise from loops involving neutrinos, making them relevant for $M_N< \mu < M_S$. We discuss their effect in great detail at the end of this section.\footnote{For the SM couplings, we will use as initial conditions their values at the top mass scale as given in Ref.~\cite{Huang:2020hdv} and integrate their beta functions at three-loop accuracy, including the most important four-loop term for the strong gauge coupling $g_s$. At $\mu=M_N$, we additionally take into account the threshold corrections given in Ref.~\cite{Brdar:2019iem}.}

Furthermore, anticipating that the potential barrier protecting the EW vacuum is linked to RG effects changing the sign of $\lambda$, we choose as RG scale $\mu$ the \textit{instability scale} $\mu_I$, defined by $\lambda(\mu_I, \eta=\mu_I)=0$. This allows us to bring Eqs.~(\ref{lambda_mu_eta})-\eqref{eq:deltalambda} to the more compact form
\begin{align}
    \lambda (\mu,\eta) \simeq - |\beta_\lambda (\mu_I)| \ln \left( \frac{\eta}{\mu_I} \right)\,.
    \label{ansatz}
\end{align}
Whether or not a barrier forms within this potential can be investigated by counting the local extrema of the potential, corresponding to solutions of the equation
\begin{align}
    0=\partial_\eta V_{\gamma}(\eta)= \eta \left[ - \frac{m_h^2}{2} + \eta^2 \left( - |\beta_\lambda| \left( \frac{1}{4} + \ln \left( \frac{\eta}{\mu_I} \right) \right)  
     + 4F \left( \frac{\eta}{M_s} \right) +  F^\prime \left( \frac{\eta}{M_s} \right) \cdot \frac{\eta}{M_s} \right)   \right]. \nonumber 
\end{align}
The number of its solutions for a given value of $m_h^2$ can be obtained by solving for $m_h^2$,
\begin{align}
    m_h^2= & 2 \eta^2 \left[  - |\beta_\lambda| \left( \frac{1}{4} + \ln \left( \frac{\eta}{\mu_I} \right) \right) + 4 F\left( \frac{\eta}{M_s} \right) +  F^\prime \left( \frac{\eta}{M_s} \right) \cdot \frac{\eta}{M_s} \right]  \label{firstterm}\\ 
     &= e^{-1/2} \mu_I^2 x^2 \left[ - 2 |\beta_\lambda| \ln (x)+ 8 F \left(z\cdot x \right) + 2 F^{\prime} \left(z\cdot x \right) z\cdot x  \right]  \\
     & \equiv e^{-1/2}  \mu_I^2 R_z (x), \label{eq:mh2bound}
\end{align}
with
\begin{align}
x\equiv e^{1/4}\, \frac{\eta}{\mu_I}\,,\qquad z\equiv e^{-1/4}\, \frac{\mu_I}{M_s} <1\,,
\end{align}
and where the last inequality holds by construction, see Eq.~(\ref{scales});  note that only the first term in Eq.~(\ref{firstterm}) depends explicitly on $\mu_I$ at the order we work in.

\begin{figure}[h!]
    \centering
    \includegraphics[width = 0.49\textwidth]{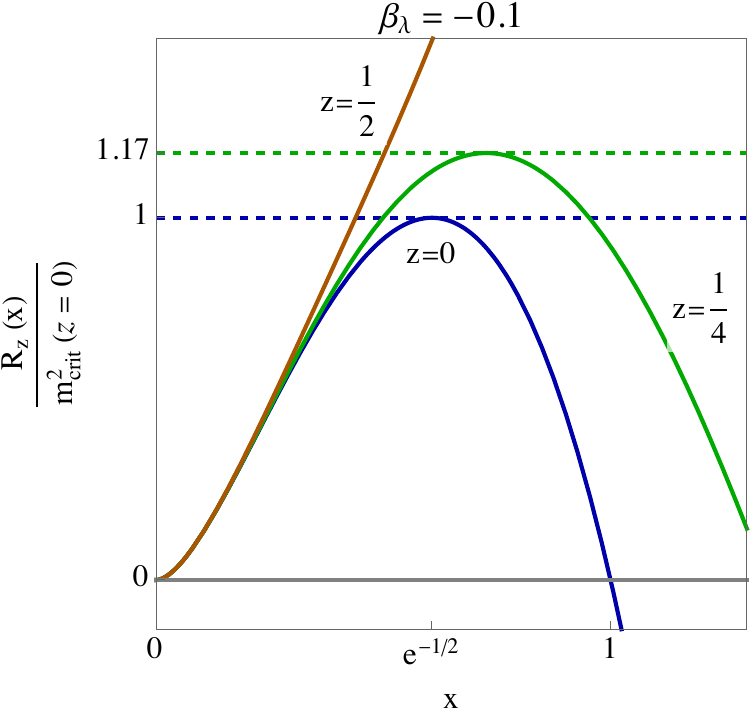}
    \includegraphics[width = 0.46\textwidth]{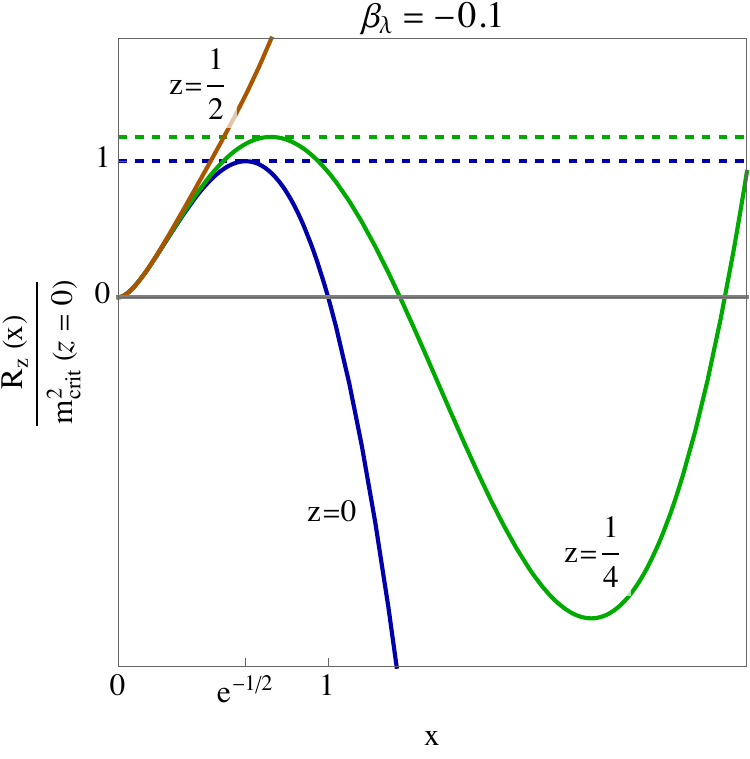}
    \caption{The function $R(x)$, with $x\equiv e^{1/4}\eta/\mu_I$, controlling the number of extrema in the effective potential $V_\gamma$ for generic scalar parameters. Different curves represent different values of the instability scale relative to $M_S$, described through the parameter $z= e^{-1/4} \mu_I/M_S$. If this parameter becomes too large, the structure of the effective potential changes, presenting the formation of a second minimum. Within the range of suitable parameters, the value of $R(x)$ agrees to leading order with the value obtained by considering only the dimension-four terms in the potential.
    }
    \label{fig:R(x)}
\end{figure}

In the limit in which the impact of the higher-dimensional polynomial terms in $R_z(x)$ can be disregarded, i.e., in the limit $F\to 0$ which holds for $\eta/M_S\to 0$, only the first term in Eq.~(\ref{firstterm}) remains, which is the well-known result in the literature (e.g. for the SM)~\cite{Buttazzo:2013uya,Khoury:2021zao}. This case has been illustrated by the line labeled $z=0$ in Fig.~\ref{fig:R(x)}. It was analyzed in detail in Ref.~\cite{Steingasser:2023ugv}, showing that the curve $R_z (x)$ with $z=0$ allows for a maximum value of the parameter $m_h$, 
\begin{align}\label{eq:bound}
    m_h^2 < m_{\rm crit}^2 \equiv \mu_I^2\,\, e^{-1/2} \,\underset{x}{\rm locmax}  \{R_z(x)\} \,,
\end{align}
where  \textit{locmax}$\{\}$ stands for the unique local maximum of the function in brackets, in this case $R$ (which is unbounded from above).

For Higgs mass values above $m_{\rm crit}$ no non-trivial solution exists, while the solution $\eta=0$ describes a global maximum. Note that $m_{\rm crit}$ is \textit{not a prediction} for the Higgs mass parameter, but indeed an \textit{upper bound}. Saturation of this bound would correspond to a potential with no barrier protecting the EW vacuum.

Let us now consider the complete equation, that is, including the impact of the terms collected in 
$F$. \cref{fig:R(x)} illustrates how its polynomial terms become increasingly relevant with larger  $z$ values. It also illustrates how, whatever the value of $M_s$, for sufficiently low values of $\eta$ (i.e.,  $x\to0$) all  $R_z(x)$ curves must converge towards that for $z=0$. For large enough values of $x$ instead, the  higher-order polynomial terms in $F$  do become relevant, causing the function $R_z$ to turn back positive. For a large range of intermediate $z$ values, this implies the existence of another solution to the equation independent of $m_h^2$, which can be identified with the true vacuum. A metastability bound appears then obeying   Eq.~(\ref{eq:bound}) above, for the given $z$ value. Following our analysis in Sec.~\ref{sec:RGV}, we can expect this behavior to be universal for the potential~$V_\gamma$, implying that the contribution of the higher-dimensional operators to Eq.~\eqref{eq:mh2bound} is generically positive and grows with $x$. As $z$ grows farther towards one, the F terms become large enough to dominate before the local maximum can form, in contradiction with our assumptions. As it can be seen in Fig. 8, this can happen even if $z<1$.

Overall, a plausible rough estimate of the requirement for  the existence of a second minimum is $\mu_I^2 \lesssim \mathcal{O}(1) |\beta_\lambda | M_S^2$, which was obtained from the analysis of a toy model~\cite{Steingasser:2023ugv}, and which sustains our initial assumption about the hierarchy of scales in Eq.~(\ref{scales}). The transition to the existence of a second minimum can itself be understood as a quantum phase transition of the potential.
 
On a practical level, the discussion above  enables us to separate the discussion of the fermionic and bosonic sectors of our model, thereby drastically simplifying the analysis: instead of performing a complete scan over the five parameters of our model -- $M_\nu, Y_\nu, M_S,\lambda_S$ and $\kappa$ -- we can first consider the neutrino sector on its own to identify regions of parameter space giving rise to a reasonably strong metastability bound. Next, and for selected benchmark points within those regions, we can perform a lifetime calculation to identify in which range of the scalar parameters the destabilizing effects of the RHNs are compensated enough so as to be in compliance with the observed age of the Universe.

\begin{figure}[h!]
    \centering
\begin{subfigure}[b]{0.4675\textwidth}
\centering
\includegraphics[width = \textwidth]{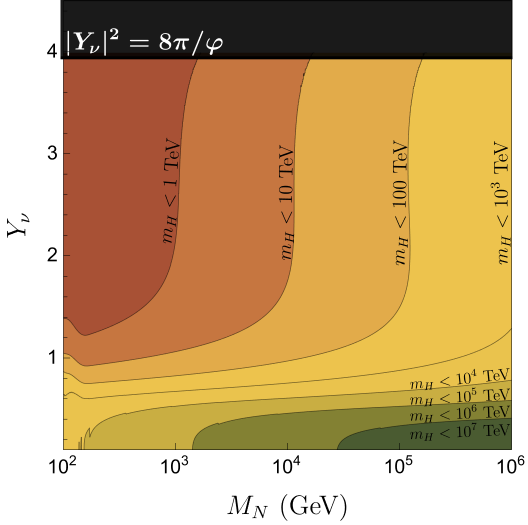}
\vspace{-0.4cm}
\end{subfigure}
\begin{subfigure}[b]{0.50\textwidth}
\includegraphics[width = \textwidth]{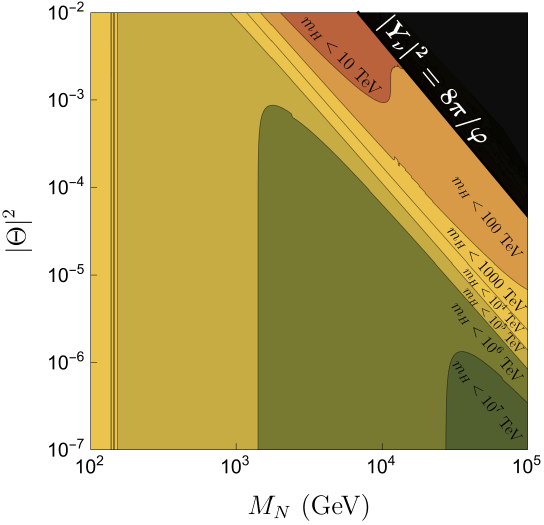}
\end{subfigure}
\caption{\textbf{Metastability bound $\bld{m_h \leq 10^k}$\,GeV} The metastability bound as given in \cref{eq:metastability-bound} as a function of the neutrino mass $M_N$ and the Yukawa coupling $Y_\nu$ in \cref{eq:majoron:yukawa-couplings}. 
As long as $M_s$ is large enough to allow the running of $\lambda(\mu)$ until $\mu = \mu_I$, the bound does not depend on the scalar sector.}\label{fig:nuRes}
\end{figure}

In summary, the metastability bound's ability to explain the observed value of the Higgs mass requires the very specific ordering of scales $m_{N} \lesssim \mu_I \lesssim M_s$, and the contribution of the heavy scalar S to the bound can be expected to be subdominant. Altogether, this means that the lowering of the metastability bound depends mainly on the details of the heavy fermionic sector, whose impact we analyse next.

\subsubsection*{The role of fermions}
The requirement of lowering the maximum possible value for the Higgs mass via the metastability bound -- Eq.~\eqref{eq:bound} -- towards $\mathcal{O}({\rm TeV})$ (so as to justify the observed value of  the Higgs mass), demands from seesaw scenarios very large values for the combinations $\text{Tr}(Y_\nu^\dagger Y_\nu)$ and $ \text{Tr}(Y_\nu^\dagger Y_\nu Y_\nu^\dagger Y_\nu)$, as well as a relatively low HNL mass $M_N$. While this low-scale scenario would contradict the canonical Type-I seesaw reasoning, there exist multiple ways to reconcile such low HNL mass values with the lightness of the observed neutrinos, such as exploiting an additional approximate $U(1)_L$ symmetry~\cite{Branco:1988ex}-\cite{Shaposhnikov:2006nn}. In these scenarios, and for the particular SPSS realization~\cite{Shaposhnikov:2006nn} used as a concrete example below, it is well-understood that the impact of the HNLs on the running of the quartic coupling $\lambda$ can effectively be described through a single parameter $|Y_\nu|$~\cite{Chauhan:2023pur}, see Eq.~(\ref{eq:majoron:yukawa-trace-def}). At the level of the relevant beta functions, this manifests as a replacement of the SM beta function $\beta_\lambda$ by 
\begin{align}
    \beta_\lambda \to & \beta_\lambda + \frac{1}{(4 \pi)^2} \left(4 \lambda |Y_\nu|^2 - 2 |Y_\nu|^4 \right) +...
\end{align}
which indeed drives $\lambda$ faster towards negative values  if $Y_\nu \gtrsim 1$. For all beta functions at the level of accuracy used in our analysis, as well as for the threshold corrections needed for the matching, see Ref.~\cite{Chauhan:2023pur}.

Our results for the metastability bound as a function of the two most relevant parameters of the full theory, $Y_\nu$ and $M_N$, are shown in the left panel of Fig.~\ref{fig:nuRes}. In the right panel of Fig.~\ref{fig:nuRes}, those same results have been projected on the parameter space $\{M_N,\,\theta_\nu \}$, where $\theta_\nu  $ is the experimentally relevant mixing angle in Eq.~(\ref{eq:majoron:mixing-angle}). For simplicity, we chose  to neglect in  these plots the (subdominant) dependence of the bound on the contributions from higher-dimensional polynomial terms  in $V_\gamma$, i.e., to neglect the $F$-dependent terms on the right-hand side of Eq.~(\ref{firstterm}). Besides allowing for a cleaner analysis of the leading-order dependence on $Y_\nu$ and $M_N$, this also makes it straightforward to relate and compare our bounds to those obtainable in models relying on different particles to lower $\mu_I$, which might rely on a different ultra violet (UV) completion to avoid insufficiently short lifetimes~\cite{Benevedes:2024tdq,Detering:2024vxs}.  

The  contour lines separating colored regions indicate the corresponding limit on the Higgs mass. The vertical shape of these contours reflects  the need for the HNLs to be lighter than the metastability bound itself, so as to be dynamical at those energies, while the horizontal shape reflects the need of large enough LHN Yukawa couplings as these constitute the portal between the LHN and the SM sector. Indeed, a stringent Higgs mass bound close to the observed $m_h$ value requires large Yukawa couplings and relatively low HNL masses, as intuitively needed for a strong impact on the RG evolution of $\lambda$. Present experimental constraints on HNL masses and mixings exclude part of the parameter space depicted, see Fig.~(\ref{fig:scalar-limits-prospects}); this will be discussed further below, see Fig.~(\ref{fig:results}). 

In conclusion, the BSM heavy fermion sector was shown to be the key to a stringent metastability bound on the Higgs mass. In turn, we will show next that the heavy BSM scalar sector is the key to ensure a viable lifetime for the Universe.

\subsection{Tunneling computation}
\label{sec:tunnelling}
In the absence of any other additional BSM physics, introducing new fermions capable of lowering the instability scale has the collateral effect of destabilizing the vacuum. The tunneling rate per unit volume can be expressed in the usual imaginary-time picture as~\cite{Coleman:1977py}
\begin{align}
    \frac{\Gamma}{V}= A \cdot e^{- S_E} \simeq \mu_S^4 \cdot e^{- S_E }\,,
    \label{eq:tunneling-rate}
\end{align}
where $\mu_S$ is an energy scale which roughly coincides with the typical field values of the so-called \textit{instanton}, while $S_E$ is the Euclidean action of this configuration~\cite{Coleman:1977py,Andreassen:2016cvx}. The instanton is a solution of the Euclidean equations of motion in imaginary time with appropriate boundary conditions. For the SM potential, the Euclidean action takes the simple form~\cite{Linde:1981zj,Isidori:2001bm,Buttazzo:2013uya}
\begin{align}\label{eq:SESM}
    S_E= \frac{8 \pi^2}{3| \lambda (\mu_S)|},
\end{align}
while the instanton scale $\mu_S$ can be identified with the RG scale where $\lambda (\mu)$ reaches its minimum~\cite{Andreassen:2017rzq}. In cases where no such minimum exists below the Planck scale, the scale $\mu_S$ becomes sensitive to gravitational effects, resulting in a value relatively close but below $M_{\rm Pl}$~\cite{Khoury:2021zao,Chauhan:2023pur}.

Introducing additional fermions to lower the instability scale $\mu_I$ also leads to significantly lower vales of $\lambda (\mu_S)$. Combining Eqs.~\eqref{eq:tunneling-rate} and~\eqref{eq:SESM}, this implies a larger decay rate, and hence, shorter lifetime of the vacuum than in the SM. Lowering the instability scale to~$\mathcal{O}$(TeV), this effect is sufficiently strong to lower the lifetime of the EW vacuum below the current age of the Universe. This scenario is therefore only in agreement with our ongoing existence in models also containing \textit{additional} new physics capable of stabilizing the vacuum. In the Majoron completion, this new physics can be identified with the additional scalar $S$ responsible for the mass of the HNLs. It is then well-understood that in the regime of interest for us a reliable calculation of the tunneling rate has to take into account the complete two-field potential in Eq.~\eqref{eq:majoron:potential}. 

\begin{figure}
    \centering
\includegraphics[width= 0.5\textwidth]{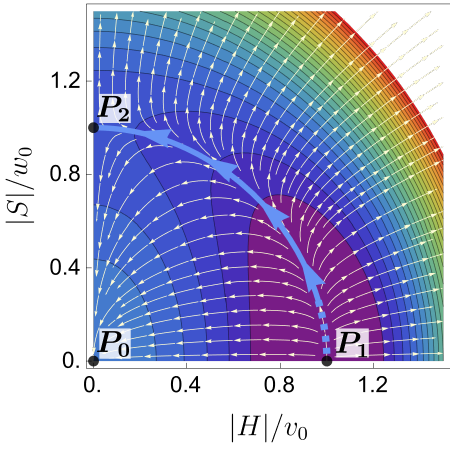}
    \caption{The potential $-V(H,S)$ determining the shape of the instanton. The white lines illustrate the steepest descend contours away from the blue trajectory $\gamma$. Eq.~\eqref{eq:shootingfull} can be understood as a point particle rolling down this potential after starting from rest, subject to a time-dependent friction. If this motion leads too far away from the center of the ridge formed by $\gamma$, the corresponding particle would ``fall off'', motivating our ansatz leading to Eq.~\eqref{eq:shootingquick}. This motion can be expected to start close to, but not exactly in $\bld{P_1}$. We visualize this through the solid blue line, with the dashed blue line representing the part of $\gamma$ not covered by it. }
    \label{fig:ridge}
\end{figure}

Hence, it is necessary to compute the Euclidean action of the two-field instanton $(H_I,S_I)$ connecting the metastable vacuum near $\bld{P_2}$ with the basin surrounding the true vacuum $\bld{P_1}$. Using the standard Ansatz of an $O(4)$-symmetric solution, i.e., $H_I(\rho )$ and $S_I(\rho )$ with $\rho^2= t^2 + \mathbf{x}^2$, the Euclidean equations of motion can be interpreted as the motion of a point particle in the inverted two-dimensional potential while subject to a time-dependent friction,
\begin{align}\label{eq:shootingfull}
    \frac{\text{d}^2}{\text{d} \rho^2}(H_I,S_I)+ \frac{3}{\rho} \frac{\text{d}}{\text{d} \rho}(H_I,S_I)= \nabla_{H,S} V(H_I,S_I).
\end{align}
We can now simplify this system of equations by recalling that the true and false vacuum are connected through a steep valley $\gamma$, which translates to a narrow ridge in the imaginary-time picture. It is now easy to see that the only path along which the ``particle'' can roll towards the true vacuum needs to be close to $\gamma$, as it would otherwise develop some runaway behavior away from the false vacuum, see Fig.~\ref{fig:ridge}. This suggests that, to leading order, we can approximate the shape of the instanton by $(H_I (\rho), S_I (\rho) = H_\gamma (\eta_I (\rho)) , S_\gamma (\eta_I (\rho)) )$, where $\eta_I (\rho)$ is the instanton in the effective potential $V_\gamma$~\cite{Buttazzo:2013uya},
\begin{align}\label{eq:shootingquick}
    \frac{\text{d}^2}{\text{d} \rho^2} \eta_I+ \frac{3}{\rho} \frac{\text{d}}{\text{d} \rho}\eta_I= \frac{\text{d}}{\text{d}\eta} V_\gamma (\eta_I).
\end{align}
The Euclidean action along this contour is then defined through
\begin{align}\label{eq:SEridge}
    S_E = \int {\rm d}^4 x \ \frac{1}{2} \dot{\eta}_I^2 + \frac{1}{2}|\nabla \eta_I|^2 + V_\gamma(\eta_I)\,.
\end{align}
On a practical level, this observation will allow us to replace the two-parameter shooting that would be necessary to solve Eq.~\eqref{eq:shootingfull} by a single-parameter shooting using Eq.~\eqref{eq:shootingquick}. 

We obtain the instanton in the potential $V_\gamma$ through the usual shooting procedure, as described, e.g., in Ref.~\cite{Benevedes:2024tdq}. When doing so, we will incorporate important loop corrections by using the RG-improved effective potential: for $M_N \lesssim \eta \lesssim M_s$, the large Yukawa couplings of the HNLs give rise to large quantum corrections, while the tree-level contribution is small in the vicinity of the instability scale: we thus replace in that regime the tree-level quartic coupling with its RG-improved effective coupling. Note that, since in that regime the Higgs mass becomes loop-suppressed relative to the remaining particles in the spectrum, this procedure allows for a consistent perturbative expansion~\cite{Gleiser:1993hf,Andreassen:2016cvx,Steingasser:2023gde}. Above $M_s$ the couplings are again dominated by their tree-level contributions, while the effects on the running due to the HNL Yukawa couplings is partially  compensated for by the portal coupling $\kappa$. Moreover, we consistently find that all relevant field values exceed $M_s$ only by factors of $\mathcal{O}(1)$. Similarly, below the scale $M_N$ the effects of the HNL become inactive, allowing for a tree-level treatment. Besides capturing important physical effects, this procedure also accelerates our shooting mechanism. Relying exclusively on the tree-level potential would instead imply the absence of a potential barrier protecting the EW vacuum, thus preventing an undershooting. 

For a given vacuum decay rate per unit volume, the lifetime of the vacuum is defined as the time after which the probability that a vacuum bubble has nucleated within the past lightcone $\mathcal{P}$ of any observer becomes of order one~\cite{Buttazzo:2013uya},
\begin{equation}
	1\sim \int_{\mathcal{P}} \text{d}^4 x \ \frac{\Gamma}{V}.
    \label{total-probability}
\end{equation}
It therefore does not only depend on the rate itself, but also on the evolution of spacetime, and in particular, the spacetime volume of the past lightcone.\footnote{A priori, the tunneling rate also depends on the temperature~\cite{Linde:1980tt,Linde:1981zj,Steingasser:2024ikl,Steingasser:2023gde}. These effects are, however, only relevant for a negligible fraction of the past lightcone, making their contribution to the total probability in Eq.~(\ref{total-probability}) negligible.} Current cosmological data suggests that for the past roughly 14 billion years, the Universe has been matter dominated. Together with the observed value of the cosmological constant, $H_0 \approx 67.4 \frac{\rm km}{\rm s \cdot Mpc}$, this implies that the volume of any observer's past lightcone is given by
\begin{equation}
    V_{\mathcal{P}}=\frac{0.15}{H_0^4}=2.2 \cdot 10^{163}(\text{GeV})^{-4}.
    \label{eq:VolumeLightcone}
\end{equation}
This then translates to a lower bound on the Euclidean action, 
\begin{equation}
    S_E>367.104 + 4 \ln \left( \frac{\mu_S}{\text{GeV}} \right),
\end{equation}
where in our case $S_E$ is given by Eq.~\eqref{eq:SEridge}. Throughout this article, we will use this condition to distinguish between feasible vacua and vacua which can be ruled out due to our ongoing existence.

For significantly longer lifetimes, the volume of the past lightcone is dominated by the contribution from the dark energy-dominated era. To leading order, this implies that the lifetime of the vacuum takes the simple form~\cite{Buttazzo:2013uya}
\begin{align}
    \tau_{\rm EW} = \frac{3 H_\Lambda^3}{4 \pi} \left( \frac{\Gamma}{V} \right)^{-1},
\end{align}
where $H_\Lambda \approx 67.4 \frac{\rm km}{\rm s \cdot Mpc}$ is the Hubble constant during the energy-dominated era.

\section{Metastability searches at colliders}
It was  argued above that the low-scale Majoron model automatically incorporates both the  fermionic and the scalar ingredients that may induce a stringent metastability bound on the Higgs mass while ensuring a viable lifetime for the Universe. We present next the existing experimental bounds on both sectors and discuss the prospects to test  metastability at future colliders.

\subsection{Neutrino sector}
\begin{figure}
 \centering
 \begin{subfigure}[b]{0.46\textwidth}
 \includegraphics[width=\textwidth]{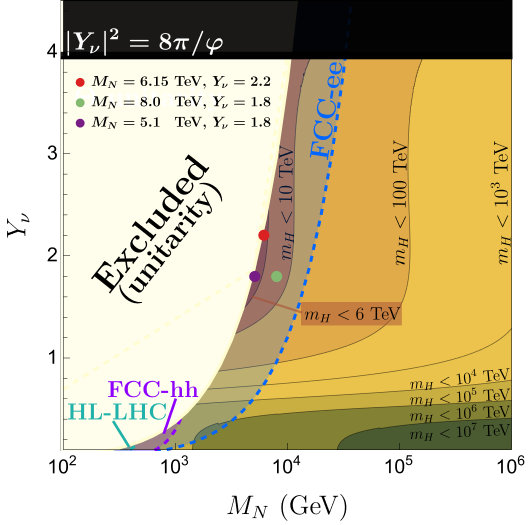}{}
 \vspace*{-0.43cm}
 \end{subfigure}
\begin{subfigure}[b]{0.49\textwidth}
 \includegraphics[width=\textwidth]{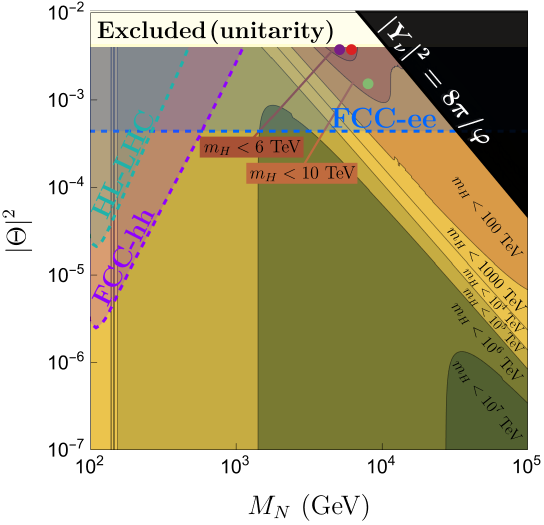}{}
 \end{subfigure}
    \caption{ HNL parameter space, showing the overlap of future collider sensitivity regions with areas where the metastability bound on the Higgs mass is stringent. The two first benchmark points in \cref{tab:benchmarkpoints} are respectively depicted as purple and red dots, both within the region $m_H\lesssim 6$\,TeV. The third point in the table is depicted in green and corresponds to a looser bound although within the $m_H\lesssim 10$\,TeV   contour.}
 \label{fig:results}
 \end{figure}

The result of superimposing the present limits and future experimental prospects for the HNL sector of the Majoron model (Fig.~\ref{fig:neutrino-limits-prospects}) over the theoretical results for the metastability bound on the Higgs mass (Fig.~\ref{fig:nuRes}) is depicted in Fig.~\ref{fig:results}, in the $\{M_N, Y_\nu\}$ parameter space and zooming into the region of $Y_\nu\sim1$.  

The shape of the contours of the metastability bounds has been discussed at the end of Sec.~(\ref{sec:RGV}). In turn, the shape of the experimentally excluded area as well as the region at experimental reach at future accelerators, included FCC-ee, is dominated by their sensitivity to unitarity constraints on the leptonic mixing matrix: the larger the HNL mass the higher the $Y_\nu$ required to induce unitarity deviations, see the discussion in Sec~(\ref{sec:experimental-limits}). The limit of the excluded region corresponds to a metastability bound of a few TeV. The figure shows the {\it solid prospects of FCC-ee to scan the entire region of strong metastability constraints above this limit}, up to Higgs mass bounds of a few tens of TeV. FCC-hh exhibits instead a modest reach to the fermionic sector of interest, limited to a small region in $Y_\nu$ values and for very stringent metastability bounds. The prospects of HL-LHC are even more restricted.
\begin{table}[h!]
    \centering
    \begin{tabular}{|c|c|c|c|c|} 
     \hline
 $\#$ & $M_N$ & $|Y_\nu|$ & $\mu_I$ & $m_{\rm crit}^2$ \\ 
 \hline
 1 & $5.1~\text{TeV}$ & 1.8 & $25.5~\text{TeV}$ & $(5.2~\text{TeV})^2$ \\  
 2 &$6.15~\text{TeV}$ & 2.2 & $19~\text{TeV}$ & $(5.6~\text{TeV})^2$ \\ 
 3 & $8.0~\text{TeV}$ & 1.8 & $39.8~\text{TeV}$ & $(7.8~\text{TeV})^2$ \\
  \hline
\end{tabular}
    \caption{The benchmark points in HNL parameter space considered in our analysis of the scalar sector.}
    \label{tab:benchmarkpoints}
\end{table}

We will focus the discussion of the final results on three benchmark points within the $\{M_N, Y_\nu\}$ region of stringent metastability bounds at reach of FCC-ee, see Table~\ref{tab:benchmarkpoints}. Two points are within the metastability bound region $m_h\leq 6$ TeV and depicted respectively in purple and red in Fig.~\ref{fig:results}. A third point -- in green -- corresponds to a somewhat higher $M_N$ value but within the metastability bound region $m_h\leq 10$ TeV.

\begin{figure}
\centering
\hspace*{0.078\textwidth}
\includegraphics[width=0.5\textwidth]{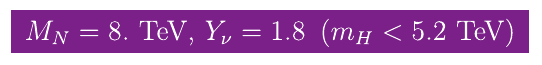}{}
\includegraphics[width=0.6\textwidth]{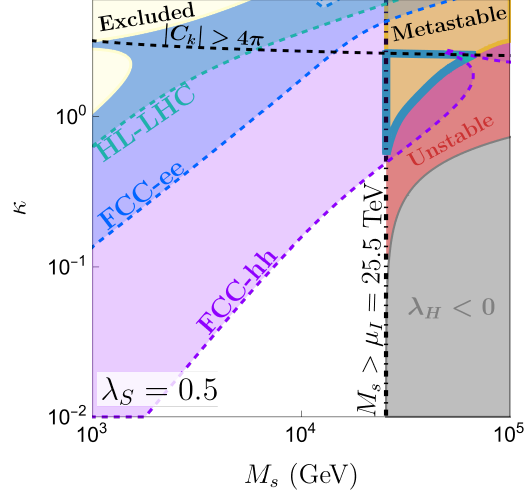}{}
\caption{Parameter space of the scalar sector in the $\{\kappa, M_s\}$ plane, for $\lambda_S = 0.5$, and for neutrino parameters   $M_N = 5.1$\,TeV, $Y_\nu = 1.8$. The latter  determine the $\mu_I$ vertical line, with the $M_s$ region on its right corresponding to a metastability bound on the Higgs mass of $5.2$ TeV in the yellow region. The desired metastability region available to the FCC-hh is the triangular-like area delimitated by a thick blue line, see text.   }
\label{fig:result-sample}
\end{figure}

\subsection{Scalar sector}

For the selected $\{M_N, Y_\nu\}$ benchmark points in Table~\ref{tab:benchmarkpoints}, we have performed a lifetime calculation to explore the scalar parameter space capable of compensating the destabilization effects of the HNLs. The destabilization manifests predominantly through the HNLs effect on the running of $\lambda$.  In the full theory, it is thus entirely captured by $\lambda_H$, which in turn is entirely determined by $\lambda$ and the remaining free parameters through the matching condition in Eq.~\eqref{matching-tree-level}, with the HNLs parameters entering when considering  one-loop effects, e.g.   $\lambda\to \lambda(\mu)$. We then scan the relevant regions of parameter space by varying $M_s$, $\kappa$ and $\lambda_S$ for each benchmark point, calculating the lifetime following the procedure described in Sec.~(\ref{sec:tunnelling}). 

In order to illustrate the analysis method, let us first focus on  benchmark point $\# 1$, in purple in Fig.~\ref{fig:results}, corresponding to $M_N=5.1$ TeV and $Y_\nu= 1.8$.  The corresponding results for the scalar sector are depicted in Fig.~\ref{fig:result-sample} in the $\{M_s,\kappa\}$ parameter space,   for $\lambda_S=0.5$. The metastable region of interest is depicted in yellow and labeled ``Metastable'', while the FCC-hh reach is depicted in light purple. The red region is excluded by the EW vacuum having a lifetime shorter than the age of the Universe. In the grey region  the ultraviolet potential is unbounded by below.

{\it The allowed region of interest which meets all requirements is the triangular-like area in 
Fig.~\ref{fig:result-sample} delimitated by a thick continuous blue line}. Per our initial assumptions, we only consider the scenario $M_s \ge \mu_I$, explaining the left vertical boundary of the allowed region. Similarly, we exclude cases with too large values of $\kappa$ to ensure perturbativity, which explains the roughly horizontal upper boundary. The non-trivial boundary to the right of the metastable area arises through the interplay of two effects: first, larger values of $\kappa$ generically stabilize the EW vacuum, since $\kappa$ is the portal through which the stabilizing effects of the heavy scalar $S$ are conveyed to the Higgs sector; second, increasing the scale $M_s$ weakens the impact of the heavy scalar on the instanton, also counteracting its stabilizing effect.

\begin{figure}
 \centering
 \begin{subfigure}{0.311\textwidth}
 \hspace*{0.055\textwidth}
 \includegraphics[width=\textwidth]{FIG-label-1.pdf}{}
 \end{subfigure}
  \begin{subfigure}{0.311\textwidth}
  \hspace*{0.06\textwidth}
 \includegraphics[width=\textwidth]{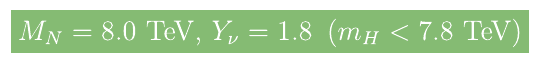}{}
 \end{subfigure}
  \begin{subfigure}{0.311\textwidth}
  \hspace*{0.06\textwidth}
 \includegraphics[width=\textwidth]{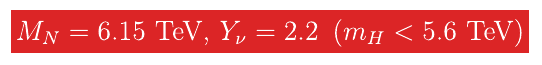}{}
 \end{subfigure}
 \begin{subfigure}[t]{0.357\textwidth}
 \includegraphics[width=\textwidth]{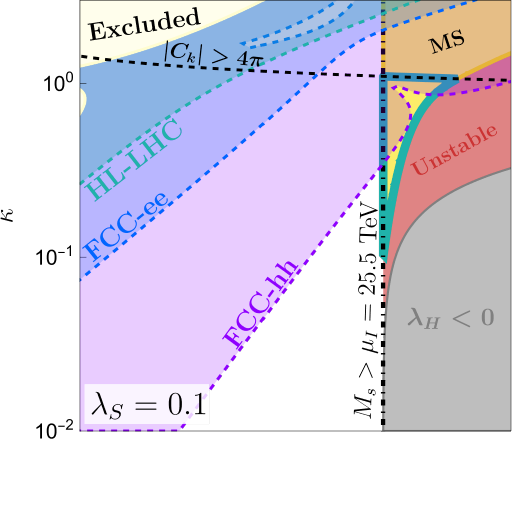}{}
 \end{subfigure}
 \begin{subfigure}[t]{0.311\textwidth}
 \vspace*{-5.5cm}
 \includegraphics[width=\textwidth]{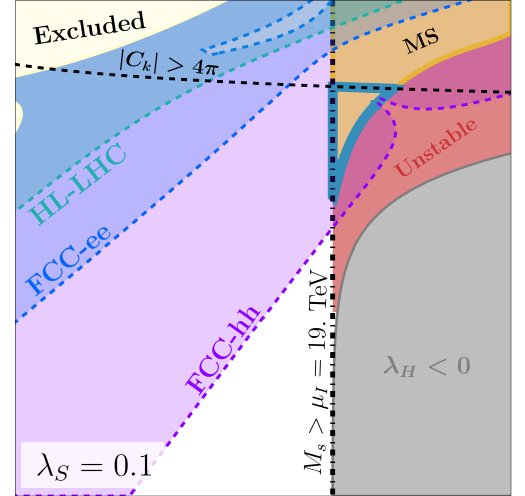}{}
 \end{subfigure}
 \begin{subfigure}[t]{0.311\textwidth}
 \vspace*{-5.5cm}
 \includegraphics[width=\textwidth]{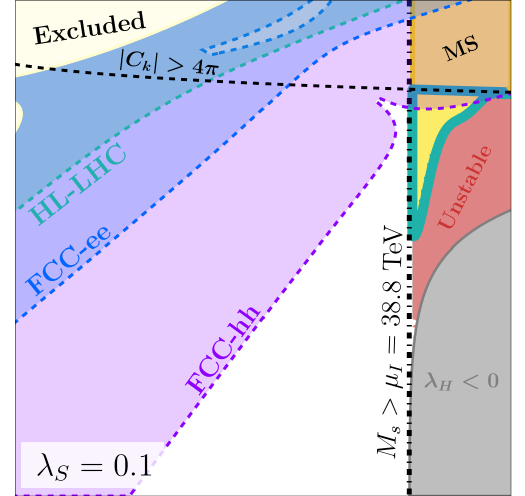}{}
 \end{subfigure}
 \begin{subfigure}[t]{0.357\textwidth}
 \includegraphics[width=\textwidth]{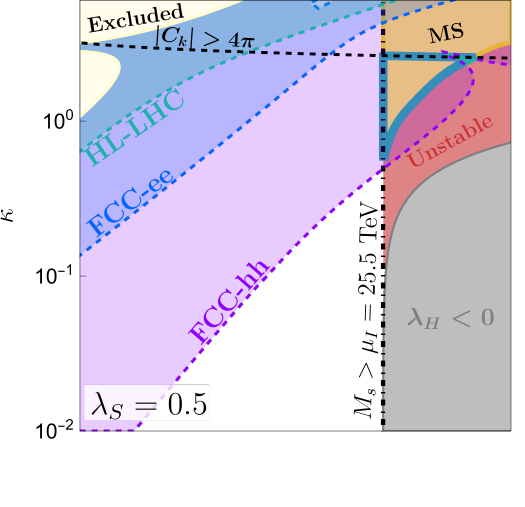}{}
 \end{subfigure}
 \begin{subfigure}[t]{0.311\textwidth}
 \vspace*{-5.5cm}
 \includegraphics[width=\textwidth]{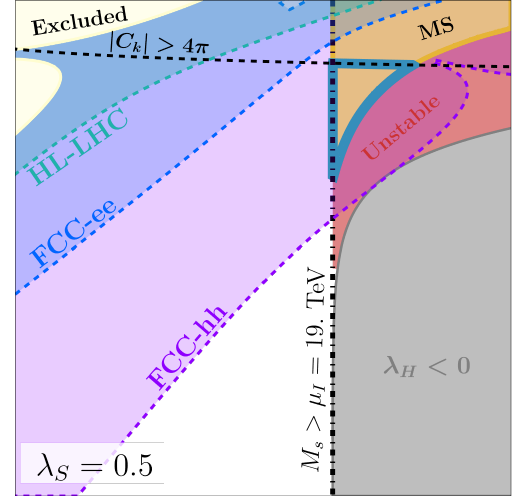}{}
 \end{subfigure}
 \begin{subfigure}[t]{0.311\textwidth}
 \vspace*{-5.5cm}
 \includegraphics[width=\textwidth]{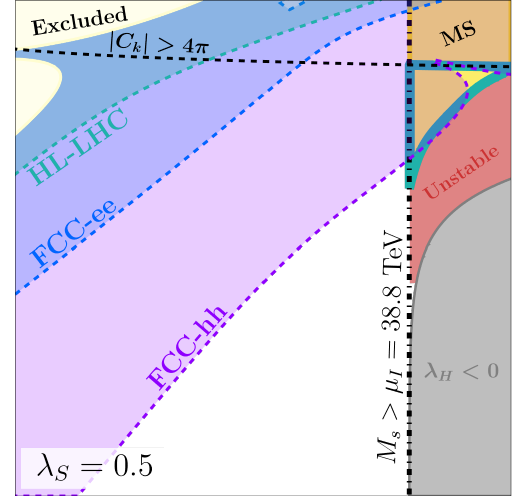}{}
 \end{subfigure}
 \begin{subfigure}[t]{0.357\textwidth}
 \includegraphics[width=\textwidth]{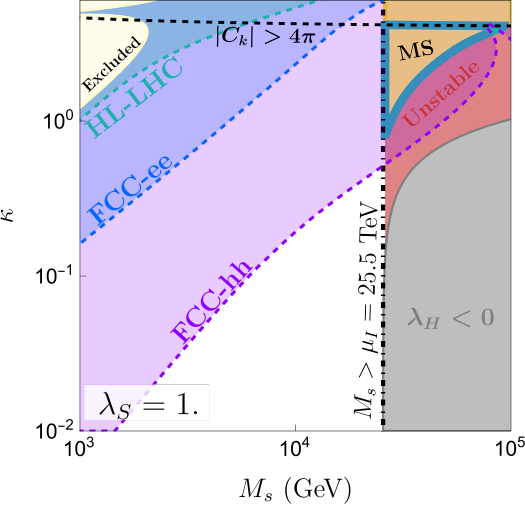}{}
 \end{subfigure}
\begin{subfigure}[t]{0.311\textwidth}
 \vspace*{-5.5cm}
 \includegraphics[width=\textwidth]{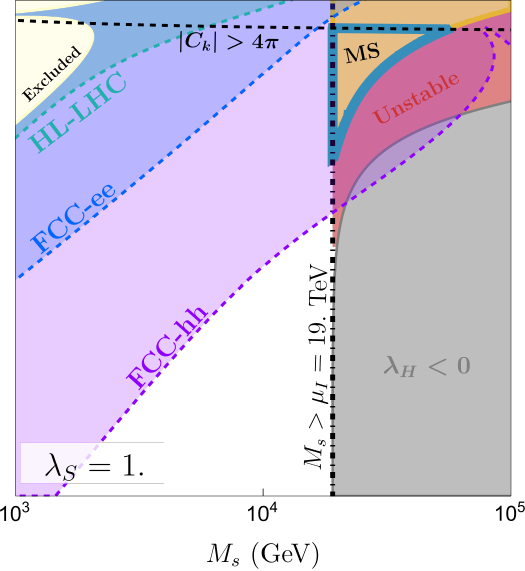}{}
 \end{subfigure}
 \begin{subfigure}[t]{0.311\textwidth}
  \vspace*{-5.5cm}
 \includegraphics[width=\textwidth]{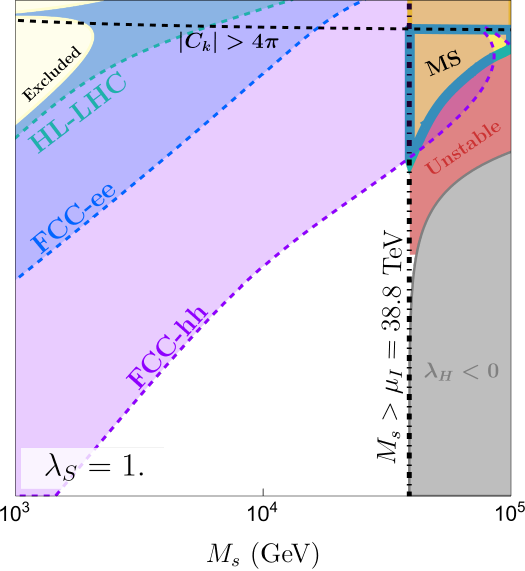}{}
 \end{subfigure}
 \caption{ Parameter space of the scalar sector in the $\{\kappa, M_s\}$ plane, for three different values of $\lambda_S$ (rows with $\lambda_S= 0.1, 0.5, 1$) and the three neutrino benchmark points in Table~\ref{tab:benchmarkpoints} (columns).  
 Description as for  Fig.~\ref{fig:result-sample}.
 The desired metastability regions available to the FCC-hh are the triangular-like areas delimitated by a thick blue line, see text. 
}
 \label{fig:results-scalar}
 \end{figure}

The effect of lowering or raising $\lambda_S$ is illustrated in  Fig.~\ref{fig:results-scalar}, together with the results for the scalar parameter space corresponding to all three benchmark points in Table~\ref{tab:benchmarkpoints}. All panels in that figure exhibit a similar pattern to that for benchmark point $\# 1$. For each point, three values of $\lambda_S$ are plotted for illustration. For given $M_s$ and $\kappa$, we find that increasing $\lambda_S$ destabilizes the EW vacuum. This effect, which is invisible in the Higgs low-energy effective theory, can also be understood in terms of the ridge potential $V_\gamma$. Recall that in our picture, the EW vacuum emerges near the saddle point $\bld{P}_2$ at $(H_{\rm EW},S_{\rm EW})=(0,M_s/(\sqrt{ 2\lambda_S}))$ -- see Eq.(\ref{P0-P1-P2}) -- while the lower-lying vacuum $\bld{P}_1$ is located at  $(H_{\rm UV},S_{\rm UV})=(\sqrt{\kappa}\,M_s/(2 \sqrt{ \lambda_S \lambda_H}),0)$~\footnote{Here, the first equality in Eq.~\eqref{eq:matchingparameters} has been applied, ensuring $m_h^2\sim 0$.}. Thus, larger values of $\lambda_S$ correspond to a smaller distance (in units of $M_s$) between these vacua in field space. In particular, in the (unphysical) limit $\lambda_S \to \infty$ they would become degenerate, implying a vanishing Euclidean action for the instanton. 

Independently of the choice of $\lambda_S$, we find that {\it the viable region of the scalar sector of the Majoron parameter space has a significant overlap with the projected range of the FCC-hh} in the region corresponding to lifetimes of the EW vacuum significantly larger than the current age of the Universe, $\tau\sim 10^{10}$~years. This is indeed consistent with the observation that the region where the  scalar sector signal is stronger at the FCC-hh is  also the region where the EW vacuum is stabilized effectively. 

\vspace{0.5cm}
In summary, FCC-ee and the FCC-hh appear to be  complementarity  as to their ability to test Higgs criticality, with FCC-ee tackling large regions of the fermion parameter space of interest and FCC-hh part of the corresponding scalar domain. For the most stringent bounds on the Higgs mass  analyzed here,  
there is a good chance that BSM signals appear both at the FCC-ee {\it and} the FCC-hh.

\newpage
\section{Conclusions}

We have presented the first  realization of the metastability bound on the Higgs mass in a concrete and complete model -- the low-scale Majoron model for neutrino masses. The Majoron model contains by construction heavy singlet neutrinos alike to those in type-I seesaw, HNLs here, and a heavy scalar, with all new scales proportional to the vev of the latter. Obtaining a strong bound requires that the new physics scale(s) lie close to the observed Higgs mass, i.e.,  $\mathcal{O}$(TeV). This in turns points to low-scale Majoron model realizations, and brings the scenario into the realm of present and foreseen colliders. 

We have first studied how the impact of the HNLs on the RG evolution of the Higgs' quartic coupling lowers the instability scale $\mu_I$ at which the latter becomes negative, while the presence of the heavy scalar is essential to stabilize the vacuum so that the lifetime of the EW vacuum is comparable or larger than the observed age of the Universe. The Majoron-model structure naturally explains a hierarchy of scales such that $M_N \lesssim \mu_I \lesssim M_s$, with all scales relatively close to one another. 

While previous analyses of the lifetime of the EW vacuum for metastability scenarios considered single-scalar potentials, the potential under study here is a two-scalar system. We have thus developed a novel approximation for the calculation of the tunneling rate in generic multi-field potentials, allowing us to take into account radiative corrections at scales where they are relevant. Our analysis shows the importance of considering ultraviolet complete models in the regime in which the metastability bound is most successful, so as to achieve a reliable calculation of the lifetime, showing that it differs substantially  from leading-order EFT approaches.

In order to assess the ability of realistic  colliders to test the paradigm of Higgs criticality, we have performed an extensive analysis of the discovery prospects at the HL-LHC and at the FCC for the low-scale Majoron model. We found that an upper bound on the Higgs mass relatively close to the observed value can be achieved through large Yukawa couplings and relatively low HNL masses, placing the interesting regions of fermionic parameter space firmly in range of existing and future colliders, in particular the FCC-ee. 

The collider impact of the additional heavy scalar has also been explored. We provide a comprehensive analysis of its testability at the FCC, and compare it to existing bounds. In order for the heavy scalar to sufficiently stabilize the vacuum, a relatively large portal coupling is needed. This translates into putative strong signals in future detectors, leading to a significant overlap of the parameter space of interest with the projected range of the FCC-hh. 

Overall, a nice complementarity appears between the FCC-ee and the FCC-hh as to their ability to test Higgs criticality, with FCC-ee tackling large regions of the fermion parameter space of interest and FCC-hh part of the corresponding scalar domain. For the most stringent bounds analyzed here (i.e., $m_h<6$ TeV, and even for $m_h< 10$ TeV), there is a good chance that BSM signals appear both at the FCC-ee {\it and} the FCC-hh. 

Two final comments are pertinent. First, the precise form used above for the low-scale Majoron model is not essential: the generic results should hold in other seesaw realizations which account for the observed light neutrino masses with low BSM scales. Second, to our knowledge this work is the first attempt to explore the sensitivity reach of future colliders to Higgs criticality using a complete BSM model. In this sense, it is a proof of concept. It remains to be seen whether the collider impact may be larger for other BSM theories leading to stricter metastability bounds on the Higgs mass.

\section*{Acknowledgments}
We thank I. Brivio, E. Fernández-Martínez, P. P. Giardino, G. Giudice, X. Marcano, M. McCullough, D. Naredo-Tuero, M. Nee and M. Ramos for illuminating discussions. Portions of this work were conducted in MIT's Center for Theoretical Physics and partially supported by the U.S. Department of Energy under Contract No.~DE-SC0012567. This project was also supported in part by the Black Hole Initiative at Harvard University, with support from the Gordon and Betty Moore Foundation and the John Templeton Foundation. The work of V.E. was supported by the Spanish MICIU through the National Program FPI-Severo Ochoa (grant number PRE2020-094281). B.G. and V. E. acknowledge as well partial financial support from the Spanish Research Agency (Agencia Estatal de Investigaci\'on) through the grant IFT Centro de Excelencia Severo Ochoa No CEX2020-001007-S, the grants PID2019-108892RB-I00 and PID2022-137127NB-I00 funded by MCIN/AEI/10.13039/501100011033/ FEDER, UE. This project has received funding/support from the European Union's Horizon 2020 research and innovation programme under the Marie Sklodowska-Curie grant agreement No.~860881-HIDDeN, and under the Marie Sklodowska-Curie Staff Exchange  grant agreement No.~101086085 -ASYMMETRY. B.G and V.E. are also indebted to MIT's Center for Theoretical Physics for hospitality while part of this work was done. The opinions expressed in this publication are those of the author(s) and do not necessarily reflect the views of these Foundations.

\newpage 

\appendix 

\section{Phase space of a 2-field potential}
\label{sec:appendix:phase-space-2-field-potential}

In \cref{sec:2-field-potential-near-criticality} the 2-field potential in the Majoron model -- see \cref{sec:majoron-completion} -- was analyzed in detail, with the aim to study its near-critical configurations. A complete classification of the possible phases in that potential according to its parametric degrees of freedom was presented, which we elaborate on in this Appendix.

The complete set of stationary points in the potential, i.e., the solutions $\bld{P}_k$ of \cref{Hminima,Sminima}, were presented in \cref{P0-P1-P2,P3} and their normalized versions $\ovbld{P}_k$ in \cref{eq:bar-P0-P1-P2,eq:bar-P3}. Clearly, the \emph{existence} condition for each stationary point $\bld{P}_k$ is that $\ovbld{P}_k$ has two real components. Are these stationary points minima, maxima or saddles? That can be determined by the Hessian $\bld{H}$ test. Let us denote
\begin{align}
    \bld{H}_k \equiv \bld{H}(H_k, S_k) 
    \equiv \left.\left(
    \begin{array}{cc}
     \displaystyle \frac{\partial^2 V}{\partial H^2}
    &\displaystyle\frac{\partial^2 V}{\partial H \partial S}\\[2mm]
     \displaystyle\frac{\partial^2 V}{\partial S \partial H}
    &\displaystyle\frac{\partial^2 V}{\partial S^2} 
    \end{array}
    \right)\right|_{H_k, S_k}\,.
\end{align}
The test establishes that a stationary point is a local minimum if $|\bld{H}_k| \equiv {\rm det}(\bld{H}_k) > 0 \wedge \bld{H}_k^{11} > 0$, a local maximum if $|\bld{H}_k| > 0 \wedge \bld{H}_k^{11} < 0$, a saddle if $|\bld{H}_k| < 0$ and is inconclusive if $|\bld{H}_k| =0$. Explicitly we have that  
\begin{align}
\begin{array}{lll}
\bld{P}_0\,: &  |\bld{H}_0| = 2\,\mu_H^2 \mu_S^2  \,,
             &  \bld{H}_0^{11} = - \mu_H^2 \,,  \\
\bld{P}_1\,: &  |\bld{H}_1| = 4\,\mu_H^2 \mu_S^2 \left[\frac{1-\Omega_H}{\Omega_H}\right]\,,
             &  \bld{H}_1^{11} = 2\,\mu_H^2 \,,    \\
\bld{P}_2\,: &  |\bld{H}_2| = 4\,\mu_H^2 \mu_S^2 \left[\frac{1-\Omega_S}{\Omega_S}\right]\,,
             &  \bld{H}_2^{11} = \mu_H^2 \left[\frac{1-\Omega_S}{\Omega_S}\right] \,, \\
\bld{P}_3\,: &  |\bld{H}_3| = 8\,\mu_H^2 \mu_S^2\,\frac{\left(\Omega_H-1\right)\left(\Omega_S-1\right)}        {\Omega_H\,\Omega_S -1} \,,
             &  \bld{H}_5^{11} = 2\,\mu_H^2 \frac{\Omega_H\left(\Omega_S-1\right)}{\Omega_H\,\Omega_S -1}  \,.\\
\end{array}
\label{eq:appendix:potential:Hessian-det+11}
\end{align}
Therefore, the parameters $\Omega_H$ and $\Omega_S$ as well as the signs of $\mu_H^2$ and $\mu_S^2$ completely determine the nature of these minima. The complete classification \emph{under the assumption of boundedness from below} -- which means applying the restrictions in \cref{eq:potential:global-stability-conditions} -- is shown in \cref{tab:potential:extrema-classification} and \cref{fig:potential:phase-space-diagrams}, while examples of all possible configurations  are displayed in \cref{fig:potential:configurations}.

\section{Kinetic canonical normalization and input scheme}
\label{sec:appendix:input-scheme}

The EFT described in \cref{sec:EFT} can be related to experiment only after an appropriate processing. Namely, it needs to be canonically normalized (mainly due to the three $d = 6$ operators already introduced in \cref{eq:SMEFT-operators}) and  an input scheme needs to be specified. All the other relevant operators for this computation can be found in \cref{tab:Warsaw-operators-SMEFT}.

\subsection*{Canonical normalization}

Two important effects have to be taken into consideration. Firstly, $\OO_H$ contributes to the Higgs potential, shifting its vacuum expectation value to 
\begin{align}
\langle H^\dagger H \rangle 
&= \frac{v_T^2}{2} \equiv \frac{v^2}{2}\left[ 1+ \frac{3}{4\,\lambda} \overline{C}_H\right] + \OO\left(\frac{1}{M_s^4}\right)\,,
&\text{where}
&& \overline{C}_H = \frac{v^2}{M_s^2} C_H\,.
\end{align}
We generalize this $\overline{C}_k$ notation to all Wilson coefficients below.
Secondly, after EWSB, now $|H| \longrightarrow \left(v_T + h\right)/\sqrt{2}$\,,
the two derivative operators generate contributions which shift the normalization of the Higgs' kinetic terms since
\begin{align}
\mathcal{O}_{H\,\square} &= 
    -\left(
    v^2 + 2\,v\,h + h^2
    \right)(\partial h)^2\, &\text{and} 
    && \mathcal{O}_{H\,D} = \frac{1}{4}\left(v^2 + 2\,v\,h + h^2\right)(\partial h)^2\,,
\end{align}
generate contributions quadratic in $\partial_\mu h$. The theory's kinetic term, 
\begin{align}
\LL &\supset \left(\frac{1}{2} - \Delta \overline{K} _H \right) (\partial h)^2\,,
&\text{with}
&&\Delta K_H \equiv \overline{C}_{H\square} - \frac{\overline{C}_{H D}}{4}\,,
\end{align}
requires finite renormalization. Several field redefinitions can be introduced to restore canonical normalization, among which adopt a non-trivial one~\cite{Hartmann:2015aia}
\begin{align}
h \longrightarrow h + \Delta \overline{K}_H \left(h + \frac{h^2}{v} + \frac{h^3}{3\,v^2}
\right)\,,
\end{align}
which contains extra terms carefully chosen to remove every momentum-dependent contribution from the to cubic and quartic terms in $\LL$. The price paid is introducing a dependence on $C_{HD}$ and $C_{H\square}$ in the Higgs' potential, which leads to the final expression   
\begin{align}
V(h) 
&= 
h^2 
\lambda v_T^2 
\left(
1 
+ 2\,\Delta \overline{K}_H 
- \frac{3}{2\lambda}\,\overline{C}_H
\right)
+ h^3 
\lambda\,v_T 
\left(
1
+ 5\,\Delta \overline{K}_H 
- \frac{5}{2\,\lambda_H}\,\overline{C}_H
\right)+\nn\\
&\hspace{7cm}   
+ h^4 
\frac{\lambda}{4} 
\left( 
1 
+ \frac{56}{3}\,\Delta \overline{K}_H
- \frac{15}{2\lambda_H}\,\bar{C}_H
\right) + \dots\,,
\end{align}
which reduces to \cref{Vhtree} if only the tree-level contributions to SMEFT coefficients are taken into account.

\subsection*{Input scheme}

There are four independent parameters in the EW sector of the SM, which can be chosen to be $\{g_1,\,g_W,\,v,\,\lambda_H\}$. Fixing them amounts to choosing an \emph{input scheme}. A commonly used input scheme, which we adopt, is the $\{\alpha_\text{e}, G_F, m_Z\}$ scheme -- we suggest Ref.\,\cite{Brivio:2020onw}, which we follow closely, for an extensive description. This scheme makes use of these three pseudo-observable quantities together with the measured mass of the Higgs, $m_h$, to fix these parameters. At  tree level, they are corrected as follows:
\begin{align}
&\alpha_{\mathrm{e}}  =\frac{1}{4 \pi} \frac{g_W^2 g_1^2}{g_W^2+g_1^2}\left[1+\Delta \alpha_{\mathrm{e}}\right]\,, 
&G_F  =\frac{1}{\sqrt{2} v_T^2}\left[1+\Delta G_F\right]\,,\\
&\Delta m_Z^2 =\frac{\bar{C}_{H D}}{2}+\frac{2 g_1 g_W}{g_1^2+g_W^2} \bar{C}_{H W B}\,, 
&\Delta m_h^2  =2 \Delta K_H-\frac{3}{2 \lambda_H} \bar{C}_H\,,
\end{align}
where
\begin{align}
&\Delta \alpha_{\mathrm{em}}  =-\frac{2 g_1 g_W}{g_1^2+g_W^2} \bar{C}_{H W B} \,,
&&\Delta G_F=(\overline{C}_{H l}^{(3)})_{11}+(\overline{C}_{H l}^{(3)})_{22}-\left(\overline{C}_{l l}\right)_{1221}\,,\\ 
&\Delta m_Z^2 =\frac{\bar{C}_{H D}}{2}+\frac{2 g_1 g_W}{g_1^2+g_W^2} \bar{C}_{H W B}\,,
&&\Delta m_h^2  =2 \Delta K_H-\frac{3}{2 \lambda_H} \bar{C}_H\,.
\end{align}
In terms of the pseudo-observable inputs $m_h^2$ and $G_F$, it is possible to rewrite the potential as 
\begin{align}
V(h) &= 
  \frac{m_h^2}{2}\,h^2
+ \frac{m_h^2G_F^{1/2}}{2^{3/4}}\,g_{hhh}\,h^3
+ \frac{m_h^2\,G_F}{4\sqrt{2}}\,g_{hhhh}\,h^4\,,
\end{align}
where we have defined
\begin{align}
g_{hhh} &= 1 - \frac{\Delta G_F}{2} + 3\,\overline{\Delta}K_H
- \frac{\sqrt{2}}{m_h^2 G_F}\overline{C}_H
\,,\\
g_{hhhh} &= 1 -\Delta G_F + 
\frac{50}{3}\,\overline{\Delta} K_H -\frac{6\sqrt{2}}{m_h^2\,G_F}\,\overline{C}_H
\,.
\end{align}

\section{Scalar field contributions to the SMEFT}
\label{sec:appendix:SMEFT-scalar}

The model in \cref{sec:majoron-completion} contains a heavy scalar field $S$, which was integrated out. In this section, we list all its one-loop contributions to the 6-dim SMEFT operators listed in \cref{tab:Warsaw-operators-SMEFT}, using the Warsaw basis~\cite{Grzadkowski:2010es}. As stated in the text, most relevant for the analysis of the potential are three operators in \cref{eq:SMEFT-operators}. At the tree level, for the Majoron model under discussion,  the contributions are 
\begin{align}
    C_H^{(0)} &= 0\,, &  C_{H\square}^{(0)} &= -\frac{\kappa^2}{4\,\lambda_S}\,,
    &
    C_{HD }^{(0)} &=0\,.
\end{align}
That is, no other operator than $C_{H\square}^{(0)}$ is generated at that level in our theory. Much more involved contributions are found at one loop. From Ref.~\cite{Jiang:2018pbd}, we find that 
\begin{align}
16 \pi^2 C_H(\mu) & = 
    16 \pi^2\left[
          C_H^{(0)}(\mu)
        + \delta C_H(\mu)
        + \left.\delta C_H(\mu)\right|_{\text {shift }}\right]+\nn\\
    &\hspace{0.6 cm}
    +\frac{\kappa^2}{9\,s_{ \theta}^2\lambda_S}\left[
    84\,\lambda^2 s_{ \theta}^2 
    - 62\,\pi\,\alpha_e\lambda
    +\left(
    60\,\pi\alpha_e\lambda 
    - 18\,s_{ \theta}^2 \lambda^2
    \right)\log\left(\frac{M_s^2}{\mu^2}\right)
    \right]\,,\\
16 \pi^2 C_{H \square}(\mu) & = 
    16 \pi^2\left[
          C_{H\square}^{(0)}(\mu)
        + \delta C_{H \square}(\mu)
        + \left.\delta C_{H \square}(\mu)\right|_{\text {shift }}\right]+\nn\\
    &\hspace{-1.7cm}
   +\frac{\kappa^2}{72\,\lambda_S}\left[
   81\,\lambda - \frac{62\,\pi\,\alpha_e}{c_{\theta}^2}
   - \frac{186\,\pi\,\alpha_e}{s_{\theta}^2}- \left(
     36\,\lambda - \frac{60\,\pi\,\alpha_e}{c_{\theta}^2}- \frac{180\,\pi\,\alpha_e}{s_{\theta}^2}
     \right)\log\frac{M_s^2}{\mu^2}
     \right]
   \,,\\
16 \pi^2 C_{H D}(\mu) &= \frac{\alpha_\text{e}\kappa^2}{9\,\lambda_S c_{\theta}^2} \left(-31+30 \log \frac{M_s^2}{\mu^2}\right)\,,
\end{align}
where the IR Higgs quartic coupling $\lambda$ was defined in Eq.~\eqref{matching-tree-level}, and we also have
\begin{align}
& \left.16 \pi^2 \delta C_H(\mu)\right|_{\text {shift }} 
=-\frac{3\,\kappa^2}{4\,\lambda_S} C_H^{(0)}\,,
&&\left.16 \pi^2 \delta C_{H \square}(\mu)\right|_{\text {shift }}  =-\frac{\kappa^2}{2\,\lambda_S} C_{H \square}^{(0)}\,,
\end{align}
as well as 
\begin{align}
16 \pi^2 \delta C_{H \square}(\mu) &=
-\left(\frac{19}{6}+\frac{9}{4}\frac{\lambda}{\lambda_S}\right)\kappa^2
- \frac{7}{12}\frac{\kappa^3}{\lambda_S}+ \frac{13}{48}\frac{\kappa^4}{\lambda_S^2}
+\nn\\
&\hspace{3cm}+\left[
\left(6+\frac{\lambda}{2\,\lambda_S}\right)\kappa^2
- \frac{\kappa^3}{\lambda_S} 
- \frac{\kappa^4}{4\,\lambda_S^2}
\right]\log\left(\frac{M_s^2}{\mu^2}\right)\,,\\
16 \pi^2 \delta C_H(\mu) &\equiv 
\frac{18\,\lambda^2}{\lambda_S}\kappa^2
+ \left(\frac{197}{12}+\frac{6\,\lambda}{\lambda_S}\right)\kappa^3 -\frac{\kappa^4}{\lambda_S}+\frac{\kappa^5}{2\,\lambda_S^2}-\frac{\kappa^6}{12\,\lambda_S^3}+\nn\\
&\hspace{3cm}-\left[
\frac{18\,\lambda^2}{\lambda_S}\kappa^2
+ \left(
\frac{63}{4}
-\frac{35}{4}\frac{\lambda}{\lambda_S}
\right)\kappa^3
\right]\log\left(\frac{M_s^2}{\mu^2}\right)\,.
\end{align}
In these expressions, $s_\theta = \sin \theta$ and $c_\theta = \cos\theta$ where $\theta$ is the Weinberg angle. The remaining 13 operators in \cref{tab:Warsaw-operators-SMEFT} are also generated at the one-loop level, but subdominant in the computation of the collider bounds in \cref{sec:experimental-limits}. However, we list them below for completeness:
\begin{align} 
16\,\pi^2 C_{HW} &= \frac{\alpha_e}{s_{ \theta}^2}\frac{\pi}{6}\frac{\kappa^2}{\lambda_S}
\,,\\
16\,\pi^2 C_{HB} &= \frac{\alpha_e}{c_{ \theta}^2}\frac{\pi}{6}\frac{\kappa^2}{\lambda_S}
\,,\\
16\,\pi^2 C_{HWB} &= \frac{\alpha_e}{c_{ \theta}s_{ \theta}}\frac{\pi}{3}\frac{\kappa^2}{\lambda_S}
\,,\\
16\,\pi^2 C_{Hu} &= \frac{\kappa^2}{\lambda_S}\left[
\frac{5}{16} y_u^\dagger y_u
-\frac{17\,\pi}{54}\frac{\alpha_e}{c_{ \theta}^2}
+\left(
\frac{\pi}{9}\frac{\alpha_e}{c_{ \theta}^2}
-\frac{y_u^\dagger y_u}{8}
\right)
\log\left(\frac{M_s^2}{\mu^2}\right)
\right]\,,\\
16\,\pi^2 C_{Hd} &= \frac{\kappa^2}{\lambda_S}\left[
\frac{17\,\pi}{108}\frac{\alpha_e}{c_{ \theta}^2}
-\frac{5}{16} y_d^\dagger y_d
+\left(
\frac{y_d^\dagger y_d}{8}
-\frac{\pi}{18}\frac{\alpha_e}{c_{ \theta}^2}
\right)
\log\left(\frac{M_s^2}{\mu^2}\right)
\right]\,,\\
16\,\pi^2 C_{He} &= \frac{\kappa^2}{\lambda_S}\left[
\frac{17\,\pi}{36}\frac{\alpha_e}{c_{ \theta}^2}
-\frac{5}{16} y_e^\dagger y_e
+\left(
\frac{y_d^\dagger y_d}{8}
-\frac{\pi}{6}\frac{\alpha_e}{c_{ \theta}^2}
\right)
\log\left(\frac{M_s^2}{\mu^2}\right)
\right]\,,\\
16\,\pi^2 C_{u H} &= \frac{\kappa^2}{\lambda_S}\frac{y_u}{72} \,\left[
45 y_u^\dagger y_u + 174\,\lambda - 124\,\pi\,\frac{\alpha_e}{s_{ \theta}^2} 
+ \right.\nn\\
&\hspace{4 cm}\left.+
\left(
120\,\pi\frac{\alpha_e}{s_{ \theta}^2} - 36\,\lambda - 54\,y_u^\dagger y_u
\right)\log\left(\frac{M_s^2}{\mu^2}\right)
\right]\,,\\
16\,\pi^2 C_{d H} &= \frac{\kappa^2}{\lambda_S}\frac{y_d}{72} \,\left[
45 y_d^\dagger y_d + 174\,\lambda - 124\,\pi\,\frac{\alpha_e}{s_{ \theta}^2} 
+ \right.\nn\\
&\hspace{4 cm}\left.+
\left(
120\,\pi\frac{\alpha_e}{s_{ \theta}^2} - 36\,\lambda - 54\,y_d^\dagger y_d
\right)\log\left(\frac{M_s^2}{\mu^2}\right)
\right]\,,\\
16\,\pi^2 C_{e H} &= \frac{\kappa^2}{\lambda_S}\frac{y_e}{72} \,\left[
45 y_e^\dagger y_e + 174\,\lambda - 124\,\pi\,\frac{\alpha_e}{s_{ \theta}^2} 
+ \right.\nn\\
&\hspace{4 cm}\left.+
\left(
120\,\pi\frac{\alpha_e}{s_{ \theta}^2} - 36\,\lambda - 54\,y_e^\dagger y_e
\right)\log\left(\frac{M_s^2}{\mu^2}\right)
\right]\,,\\
16\,\pi^2 C_{Hq}^{(1)} &=
\frac{1}{864} 
\frac{\kappa^2}{\lambda_S} \,
\left[
135\left(y_d y_d^\dagger - y_u y_u^\dagger\right)
-68\,\pi\frac{\alpha_e}{c_{ \theta}^2}
+ \right.\nn\\
&\hspace{4 cm}\left.+
\left(
24\,\pi\,\frac{\alpha_e}{c_{ \theta}^2} 
+ 54\left(y_u y_u^\dagger - y_d y_d^\dagger\right)\right)\log\left(\frac{M_s^2}{\mu^2}\right)
\right]\,,\\
16\,\pi^2 C_{Hq}^{(3)} &=
\frac{1}{284} 
\frac{\kappa^2}{\lambda_S} \,
\left[
45\left(y_u y_u^\dagger+y_d y_d^\dagger  \right)
-68\,\pi\frac{\alpha_e}{s_{ \theta}^2}
+ \right.\nn\\
&\hspace{4 cm}\left.+
\left(
24\,\pi\,\frac{\alpha_e}{c_{ \theta}^2} 
- 18\left(y_u y_u^\dagger + y_d y_d^\dagger\right)\right)\log\left(\frac{M_s^2}{\mu^2}\right)
\right]\,,\\
16\,\pi^2 C_{Hq}^{(1)} &=
\frac{1}{288} 
\frac{\kappa^2}{\lambda_S} \,
\left[
45\,y_e y_e^\dagger
+68\,\pi\frac{\alpha_e}{s_{ \theta}^2}
- 
\left(
24\,\pi\,\frac{\alpha_e}{s_{ \theta}^2} 
+ 18\,y_e y_e^\dagger\right)\log\left(\frac{M_s^2}{\mu^2}\right)
\right]\,,\\
16\,\pi^2 C_{Hq}^{(3)} &=
\frac{1}{288} 
\frac{\kappa^2}{\lambda_S} \,
\left[
45\,y_e y_e^\dagger
-68\,\pi\frac{\alpha_e}{c_{ \theta}^2}
+
\left(
24\,\pi\,\frac{\alpha_e}{c_{\theta}^2} 
- 18\,y_e y_e^\dagger\right)\log\left(\frac{M_s^2}{\mu^2}\right)
\right]\,.
\end{align}

\bibliographystyle{utphysMOD}
\bibliography{Bibliography}

\end{document}